\documentstyle[aps,preprint,epsf,eqsecnum,tighten]{revtex}

\renewcommand{\d}{{\rm d}}
\newcommand{\e}{{\rm e}}
\newcommand{\I}{{\rm i}}

\begin{document}

\draft
\preprint{nucl-th/9612059}

\title{Electroproduction of $\phi$ meson from proton \\
       and the strangeness in the nucleon}

\author{Alexander I. Titov%
\thanks{
   Permanent address:
   Bogoliubov Theoretical Laboratory, Joint Institute for Nuclear
   Research, 141980 Dubna, Russia.},
	Shin Nan Yang
and
	Yongseok Oh%
\thanks{
   Present address:
   Institut f\"ur Theoretische Physik, Physik Department,
   Technische Universit\"at M\"unchen, James--Franck--Stra{\ss}e,
   D-85747 Garching, Germany.}}

\address{Department of Physics, National Taiwan University,
         Taipei, Taiwan 10617, Republic of China}


\maketitle

\begin{abstract}
We analyze the process $e \, p \to e\, p\, \phi$ near the threshold
within $uud$-$s\bar s$ cluster model as a probe of the strangeness
content of proton. Our consideration is based on the relativistic
harmonic oscillator quark model which takes into account the
Lorentz-contraction effect of the hadron wavefunctions. We find that
the knockout mechanisms are comparable to the diffractive production
of vector-meson-dominance model when only (3--5)\% strange quark
admixture is assumed, which should be compared with (10--20)\% of the
nonrelativistic quark model prediction. The cross sections of the
$uud$- and $s\bar s$-knockout processes have qualitatively different
dependence on the four-momentum transfer squared to the proton and
may be distinguished experimentally. We also briefly discuss a way to
determine the strangeness content of proton in 5-quark cluster model.
\end{abstract}



\section{Introduction}

Conventional phenomenological quark models widely used for describing
low-energy properties of baryons treat nucleon as consisting of only
up and down quarks; intrinsically, there is no strange quark in the
nucleon. Considering the success of the constituent quark models
\cite{Close}, it may come as a surprise when some recent measurements
and theoretical analyses indicate a possible existence of a significant
strange quark content in the nucleon.

For example, analyses of the sigma term \cite{sigmaterm} in
pion-nucleon scattering suggest that about one third of the rest
mass of the proton comes from $s\bar s$ pairs inside the proton.
The EMC measurement of the proton spin structure functions in
deep-inelastic muon scattering \cite{EMC,SMC} has been interpreted
as an indication of the strange quark sea $s\bar s$ strongly polarized
opposite to the nucleon spin, leading to the conclusion that the
total quark spin contributes little to the total spin of the proton.
A similar conclusion%
\footnote{For other interpretations of these experiments, see, for
instance, Ref. \cite{ASJL}.}
has been drawn from the BNL elastic neutrino-proton scattering \cite{N1,N2}.
This has stimulated a set of new experimental proposals \cite{newexp} to
measure the neutral weak form factors of the nucleon which might be
sensitive to the strange quarks of the nucleon.

One of the intriguing idea associated with a direct probe of the
strangeness content of proton is to use $\phi$ meson production
from proton \cite{HKPW,HKW,EGK,EK}. Since $\phi$ meson is
nearly 100\% $s \bar s$-state because of the $\omega$-$\phi$ ideal
mixing, its coupling to the proton is suppressed by the OZI rule.
The main idea of this proposal, called OZI ``evasion'' process, is to
determine the amount of the $s \bar s$-admixture of nucleon, if any,
by isolating its contribution in $\phi$ production processes. For
example, $s \bar s$ admixture in the proton wavefunction allows
contributions from ``shakeout'' and ``rearrangement'' diagrams in
$\phi$ production from $p \bar p$ and $pp$ collisions \cite{EK}.
Another idea is to use $\phi$-meson lepto- and electroproduction from
proton target as advocated by Henley {\em et al.\/} \cite{HKPW,HKW}.
In this case, in addition to the diffractive production mechanism of
the vector-meson--dominance model (VDM) we have contributions from the
direct ``knockout'' mechanism.

If we consider $s \bar s$ admixture of proton, we can parameterize
the Fock-space decomposition of the proton wavefunction as
\begin{equation}
|p \rangle = A_0 | uud \rangle + \sum_X A_X | uudX \rangle
+ \sum_X B_X | uuds \bar s X \rangle,
\label{Fockp}
\end{equation}
where $X$ denotes any combination of gluons and light quark pairs of
$u$ and $d$ quarks. Analyses of the $\phi$ production experiments can
give estimates on $B_X$. For instance, investigation of $p \bar p$
annihilation by connected quark line diagrams estimates that the
$s \bar s$ sea quark contribution to the proton wavefunction is
between 1\% and 19\% \cite{EKKS}. In Ref. \cite{HKW}, by using the
$\phi$ electroproduction its upper bound is estimated to be about
10-20\%. To obtain this estimation, the authors used nonrelativistic
quark model (NRQM) and calculated the cross section of the
$s \bar s$-knockout process. Because of the paucity of experimental
data \cite{oldexpt,VDM}, it was compared with the VDM predictions%
\footnote{
The recent ZEUS experiment \cite{ZEUS} was done at very high energy,
and is beyond the applicability of this work.
}.
However, as was pointed out by our previous publication \cite{TOY94}
where a preliminary result was reported, the knockout contributions
are closely related to the hadron form factors and in the considered
kinematical region of $\phi$ production, the minimum value of
$\bbox{q}^2$ is about 3.6 GeV$^2$, where $\bbox{q}$ is the
three-momentum transferred to the hadron system. So it is clear that
the value of the momentum transfer in this process is too large to
use the NRQM because its predictions on hadron form factors are in
poor agreement with experiment at $\bbox{q}^2 \geq 0.3$ GeV$^2$.

In this paper, we improve the work of Ref. \cite{HKW} by including
relativistic effects based on the relativistic harmonic oscillator
model (RHOM) \cite{YM,FKN,FKR,Lipes,KN,KNO,BOLT} which describes
successfully the proton form factors in a wide range of $\bbox{q}^2$.
We also carry out the calculations for the $uud$-knockout and its
interference with the $s \bar s$-knockout which were argued to be
suppressed and left out in Ref. \cite{HKW} as well as for the
$s \bar s$-knockout. The calculations are done both in NRQM and in
RHOM for a comparison. As in Ref. \cite{HKW}, we will compare the
cross sections of the knockout processes with the VDM predictions.
However, this does {\it not} mean that the knockout mechanism should
replace the VDM mechanism. The latter is present as a background of
the knockout mechanism and our purpose is to determine a theoretical
upper bound of $B_X^2$ using $\phi$ electroproduction process.

This paper is organized as follows. In the next section, we briefly
review the general structure of the knockout and VDM differential
cross sections and introduce the kinematical variables for $\phi$
meson electroproduction. Then in Sect. III we discuss the proton
wavefunction of 5-quark cluster model. Section IV is devoted to the
evaluation of the knockout process matrix elements within the
non-relativistic harmonic oscillator quark model. In Sect. V we
perform the calculations based on the RHOM which provides an explanation
of the dipole-like $Q^2$ dependence of the elastic nucleon form factor.
This model, though probably it has not underlying physical significance,
has the pleasant feature that basically all quantities of interest can
be worked out analytically, and, in many cases, it allows understanding
of the qualitative picture of the reaction. The nontrivial role of the
relativistic Lorentz-contraction effect is also discussed. In Sec. VI
we briefly discuss a way to extract out the strangeness content of
proton in an extended quark model. Section  VII contains a summary and
some details in the calculation are given in Appendix.


\section{Kinematics and cross sections}

\begin{figure}
\vskip 0.7cm
\centerline{\epsfxsize=0.5\hsize \epsffile{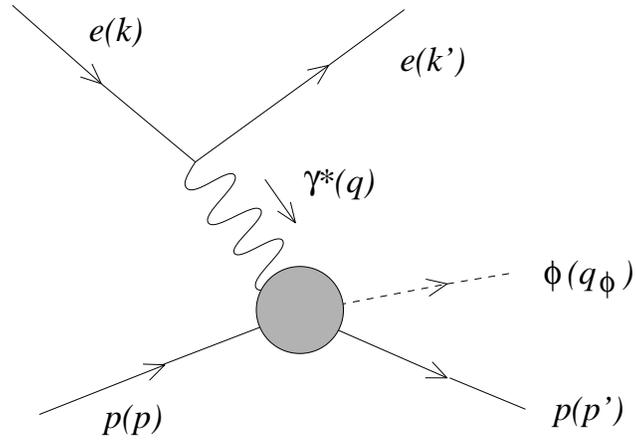}}
\vskip 1.2cm
\caption{Kinematics of the electroproduction of $\phi$ meson from 
proton, $e\,p \to e\,p\,\phi$.}
\label{phipr}
\end{figure}

The one-photon exchange diagram for $\phi$ electroproduction is shown
in Fig.~\ref{phipr}. The four momenta of the initial electron and proton,
final electron and proton, the produced $\phi$ meson, and the virtual
photon are denoted by $k$, $p$, $k'$, $p'$, $q_{\phi}$, and $q$,
respectively. In the laboratory frame, we write $k \equiv ( E_e,
\bbox{k})$, $k' \equiv ( E_e',\bbox{k}')$, $p \equiv (E_p, \bbox{p}) =
(M_N,\bbox{0})$, where $M_N$ is the nucleon mass, $p' \equiv (E_p',
\bbox{p}')$, $q_\phi \equiv (\omega_\phi, \bbox{q}_\phi)$, and $q \equiv
(\nu, \bbox{q})$, respectively. The electron scattering angle $\theta$
is defined by $\cos\theta = \bbox{k} \cdot \bbox{k}'/ \left| \bbox{k}
\right| \left| \bbox{k}' \right|$. We also denote the electron mass and
$\phi$ mass by $M_e$ and $M_\phi$, respectively. The other invariant
kinematical variables are $\nu \equiv p \cdot q / M_N = E_e - E_e'$, the
minus of photon mass squared $Q^2 \equiv -q^2$, the four-momentum transfer
squared to the proton $t \equiv (p -p')^2$, the proton--virtual-photon
center-of-mass energy $W^2 = ( p + q )^2$, and the total energy squared
in the CM system $s=(p+k)^2$. We also use dimensionless invariant
variables $\eta$, $y$ and $z$ defined as
\begin{eqnarray}
\eta = \frac{Q^2}{4 M_N^2}, \quad
y = \frac{-t}{4 M_N^2}, \quad
z = \frac{W^2 - M_N^2}{4 M_N^2}.
\end{eqnarray}
So, in the laboratory frame we have
\begin{eqnarray}
&& \bbox{q}^2 = 4 M_N^2 [ (\eta + z)^2 + \eta ] , \quad
\nu = 2 M_N (\eta + z), \nonumber\\
&& E_p' = M_N + 2 M_N y, \quad
\omega_{\phi} = \nu - 2 M_N y, \quad s \simeq M_N (2 E_e + M_N).
\end{eqnarray}

\begin{figure}
\centerline{\epsfxsize=0.45\hsize \epsffile{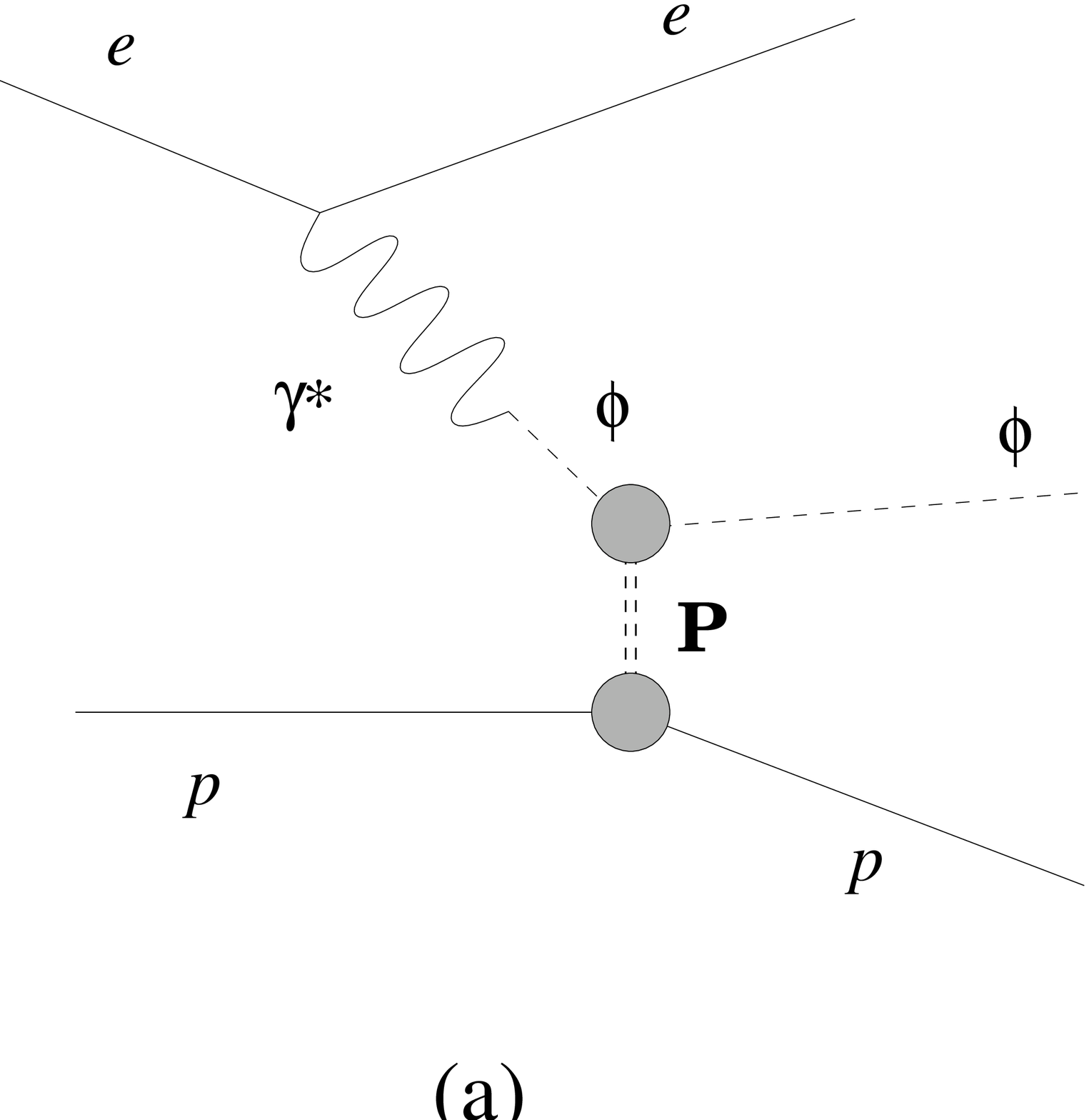}}
\centerline{\epsfxsize=0.40\hsize \epsffile{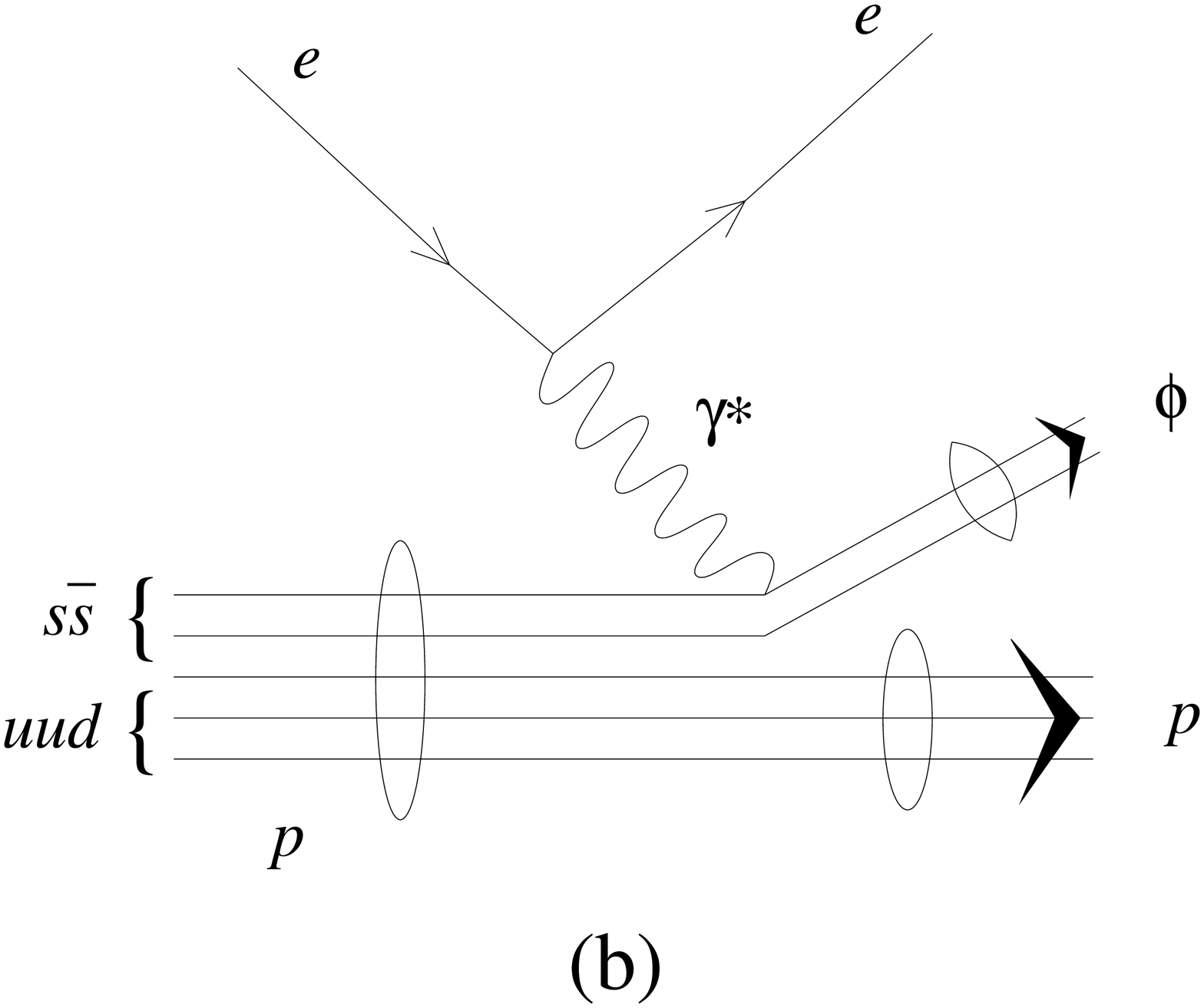}
\hfill \epsfxsize=0.40\hsize \epsffile{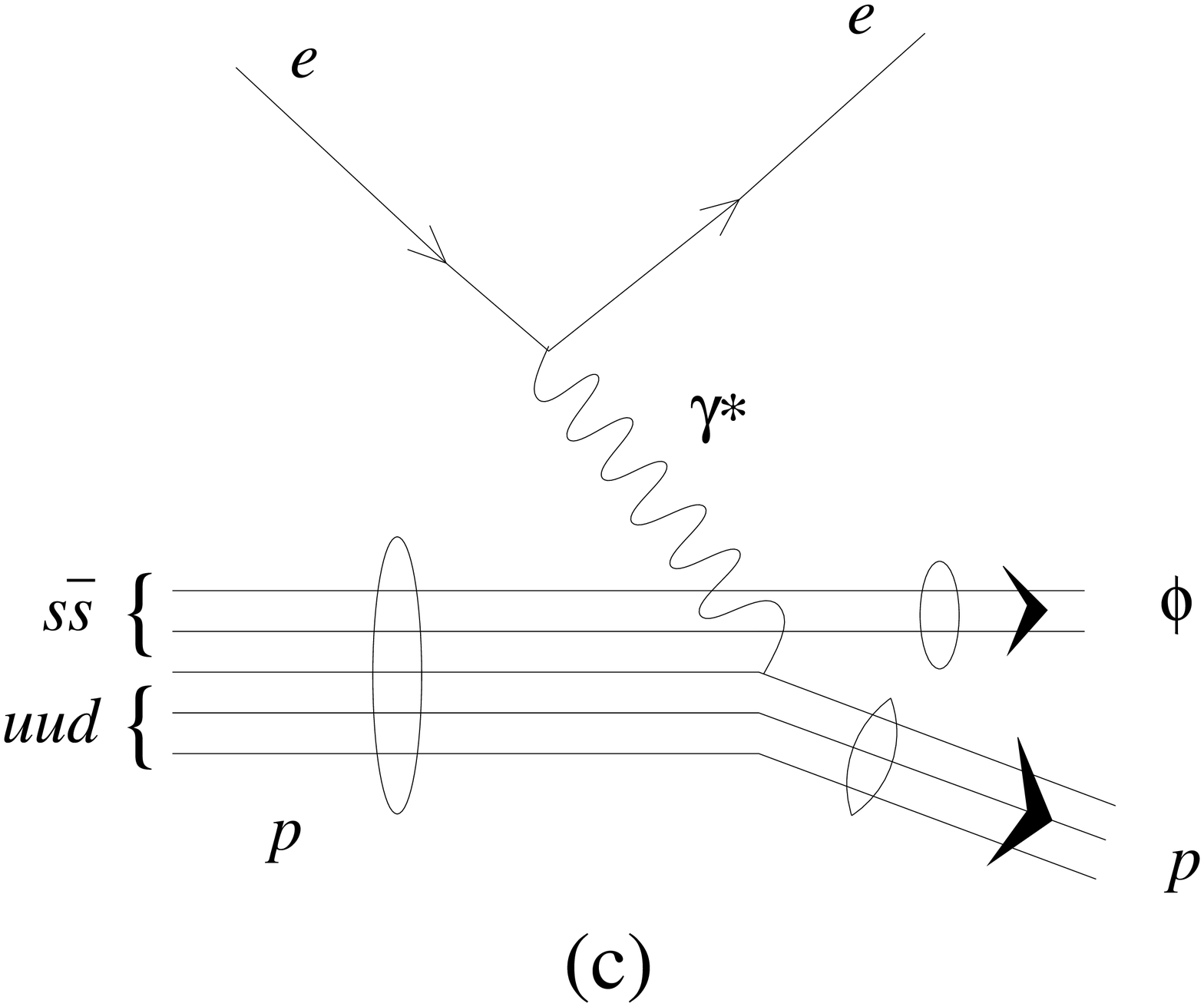}}
\vskip 2.5cm
\caption{
 (a) Diffractive $\phi$ meson production within the
     vector-meson-dominance model by means of Pomeron exchange;
 (b) $s\bar s$-knockout and (c) $uud$-knockout contributions to $\phi$
     meson electroproduction.}
\label{pros1}
\end{figure}

In terms of the conventional $T$-matrix elements $T_{fi}$, the
differential cross section is given as
\begin{mathletters}
\label{sigma}
\begin{eqnarray}
\d \sigma_{fi} = (2\pi)^4 \delta(p+q-p'-q_{\phi})
\frac{2 E_p E_e}{\sqrt{\lambda(s,M_e^2,M_N^2)}}
{\overline {\vert T_{fi}\vert^2}}
\frac{\d\bbox{k}'}{(2\pi)^3} \frac{\d\bbox{p}'}{(2\pi)^3}
\frac{\d\bbox{q}_\phi}{(2\pi)^3},
\end{eqnarray}
and
\begin{equation}
\overline {\vert T_{fi}\vert^2} =
\frac14 \sum\limits_{{m_i,m_f,m_{\phi},}\atop{m_e,m_{e'}}}
\vert T_{fi}\vert^2 ,
\end{equation}
where $m_{i,f}$ ($m_{e,e'}$) and
$m_{\phi}$ are the spin projections of the incoming and outgoing
proton (electron) and outgoing $\phi$ meson, respectively, and
\end{mathletters}
\begin{equation}
\lambda (x,y,z) = x^2 + y^2 + z^2 - 2 (xy + yz + zx).
\label{lambda}
\end{equation}
Upon integrating Eq. (\ref{sigma}) over non-fixed kinematical variables
we find the triple-differential cross section of the $\phi$
electroproduction in the laboratory frame in the form of
\begin{mathletters}
\label{CS}
\begin{eqnarray}
\frac{{\d}^3 \sigma}{\d W \d Q^2 \d t}=
\frac{W E_e' E_p' \omega_\phi}{4 M_N^2 |\bbox{k}| |\bbox{q}|}
\frac{1}{(2\pi)^3}
\overline {\overline {\vert T_{fi}\vert^2 }},
\end{eqnarray}
where
\begin{equation}
\overline{\overline {\vert T_{fi}\vert^2 }}
=\int \overline {\vert T_{fi}\vert^2 }
\frac{\d \varphi_{p'}}{2\pi} \frac{\d \varphi_{q}}{2\pi},
\label{barbarT}
\end{equation}
and $\varphi_{p',q}$ are the corresponding azimuthal angles,
respectively.

The tree diagrams that could contribute to $\phi$ electroproduction are
illustrated in Fig.~\ref{pros1}. In Fig. \ref{pros1}(a), the virtual
photon turns into the $\phi$ meson and then scatters diffractively with
the proton through the exchange of a Pomeron. This VDM of diffractive
production has been widely used to describe vector-meson photo- and
electro-productions. It generally reproduces well the $Q^2$ dependence
of the cross sections for fixed $W$, but is not successful to account for
the experimentally observed rapid decrease in the cross sections with
increasing $W$.
The double differential cross section predicted by the VDM is \cite{VDM}
\end{mathletters}
\begin{equation}
\frac{\d^2\sigma_{\rm dif}}{\d W \d Q^2} = (2\pi) \Gamma_W(Q^2,W)
\sigma_{\rm dif}(Q^2,W),
\end{equation}
where
\begin{equation}
\Gamma_W (Q^2,W) = \frac{e^2}{32\pi^3} \frac{W}{M_N E_e^2}
\frac{W^2 - M_N^2}{M_N Q^2} \frac{1}{1-\epsilon},
\end{equation}
which is related
to the flux of transverse virtual photons in the laboratory frame
$(2\pi)\Gamma_T(\theta, E_e')$ (for fixed $\theta$ and $E_e'$) by
$\Gamma_W = J \Gamma_T$, with $J$ being the Jacobian
$J(\cos\theta,E_e';Q^2,W) = W/(2 M_N E_e E_e')$. Here,
$\epsilon$ is the virtual-photon polarization parameter
\begin{equation}
\epsilon = \frac{1}{\{1+2[(Q^2+\nu^2)/Q^2]\tan^2(\theta/2)\}}.
\end{equation}

As in Refs. \cite{HKW,VDM}, we will work with the cross section
$\sigma(Q^2,W)$, which is predicted by VDM as \cite{VDM}
\begin{equation}
\sigma_{\rm dif} (Q^2,W) = \frac{\sigma_\phi(0,W)}{(1+Q^2/M_\phi^2)^2}
\frac{p_\gamma^*(0)}{p_\gamma^*(Q^2)} ( 1 + \epsilon R_\phi)
\exp( -b_\phi | t_{\rm max}(Q^2) - t_{\rm max}(0) |),
\label{sig_dif}
\end{equation}
where $\sigma_\phi(0,W)$ is the photoproduction cross section. The
range of $t$ ($t_{\rm min} < t < t_{\rm  max}$) can be obtained from
$\cos\theta_{qp'} = ( \bbox{p}^{\prime 2} + \bbox{q}^2 - \bbox{q}_{\phi}^2
) / ( 2|\bbox{p}'| |\bbox{q}| ) = \pm 1$. The first part of (\ref{sig_dif})
represents the photoproduction cross section extrapolated to $Q^2$ by
the square of the $\phi$ propagator. The second represents a correction
to the virtual photon flux where $p_\gamma^*$ is the virtual photon
momentum in $\gamma^*p$ CM frame. Explicitly it is written as \cite{FS}
\begin{equation}
\frac{p_\gamma^*(0)}{p_\gamma^*(Q^2)} = \sqrt{ \frac{(W-M_N)^2 (W+M_N)^2}
{[(W-M_N)^2+Q^2][(W+M_N)^2+Q^2]}},
\end{equation}
which can be approximated to unity in the large $W$ limit. This term
is a measure of the ambiguity in the model predictions \cite{VDM,FS},
and comes from a choice made in the definition of the transverse photon
flux. The $(1 + \epsilon R_\phi)$ term corrects the cross section for the
longitudinal component which is missing at $Q^2=0$, and the exponential
factor corrects for the fact that for a given $W$ the physical range of
$t$ is smaller when $Q^2>0$ than its range at $Q^2=0$. We fix the
parameters following Refs. \cite{oldexpt,VDM} as $b_\phi = 3.46$ GeV$^2$,
$R_\phi = \xi^2 Q^2 / M_\phi^2$ with $\xi^2=0.33$, and $\sigma_\phi(0,W)
= 0.22$ $\mu$b, which is fitted for $W = 2.1$ GeV.

The $t$-dependence of the cross section can be obtained by assuming
$\exp(b_\phi t)$ dependence. This gives
\begin{equation}
\sigma_{\rm dif} (Q^2,W,t) = \sigma_{\rm dif} (Q^2,W) \, b_\phi \,
\e^{b_\phi[t-t_{\rm max}(Q^2)]},
\end{equation}
provided that $\exp(b_\phi t_{\rm min})$ is negligible \cite{HKW}.

\begin{figure}
\centerline{\epsfxsize=0.32\hsize \epsffile{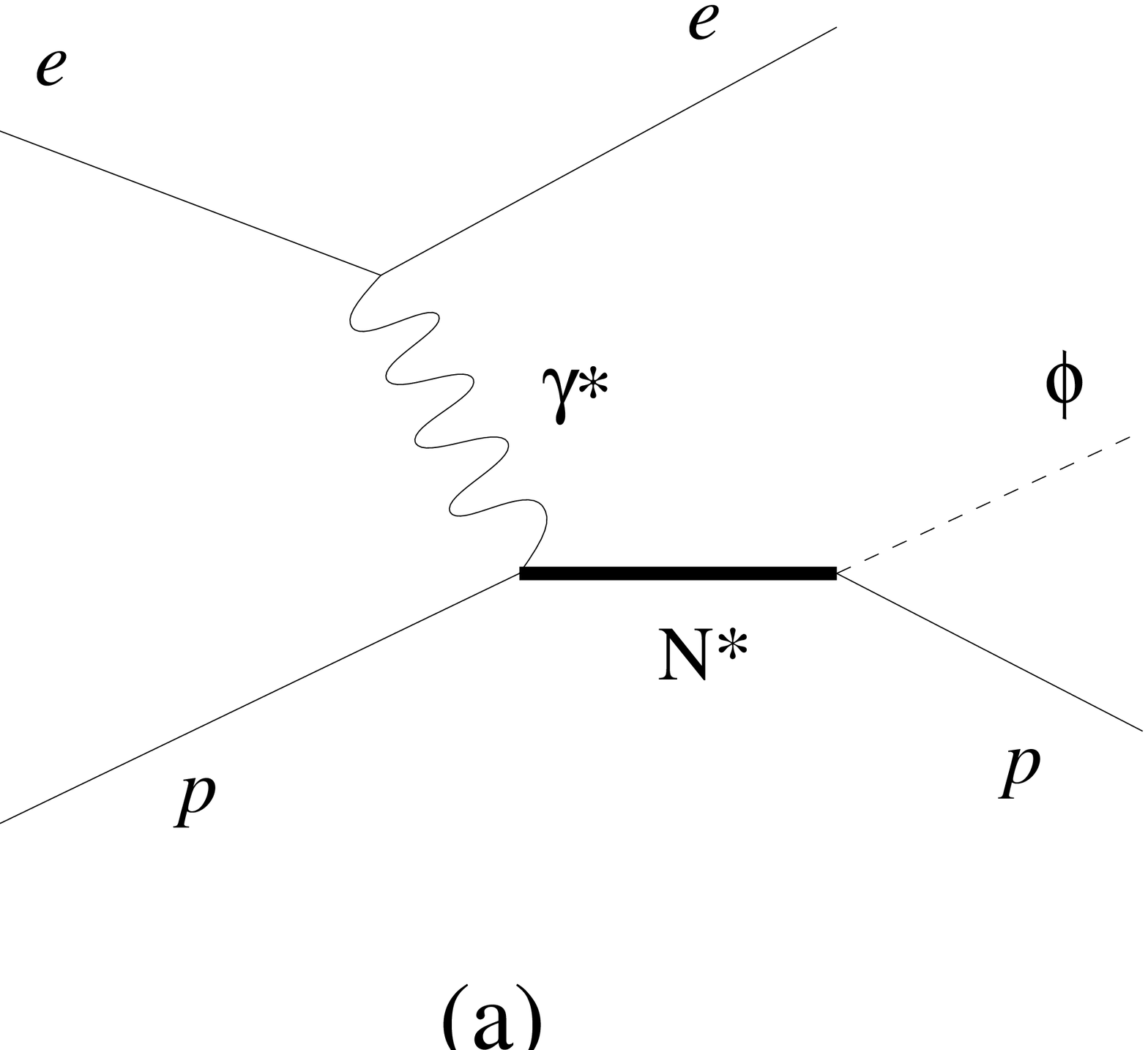}
\hfill \epsfxsize=0.32\hsize \epsffile{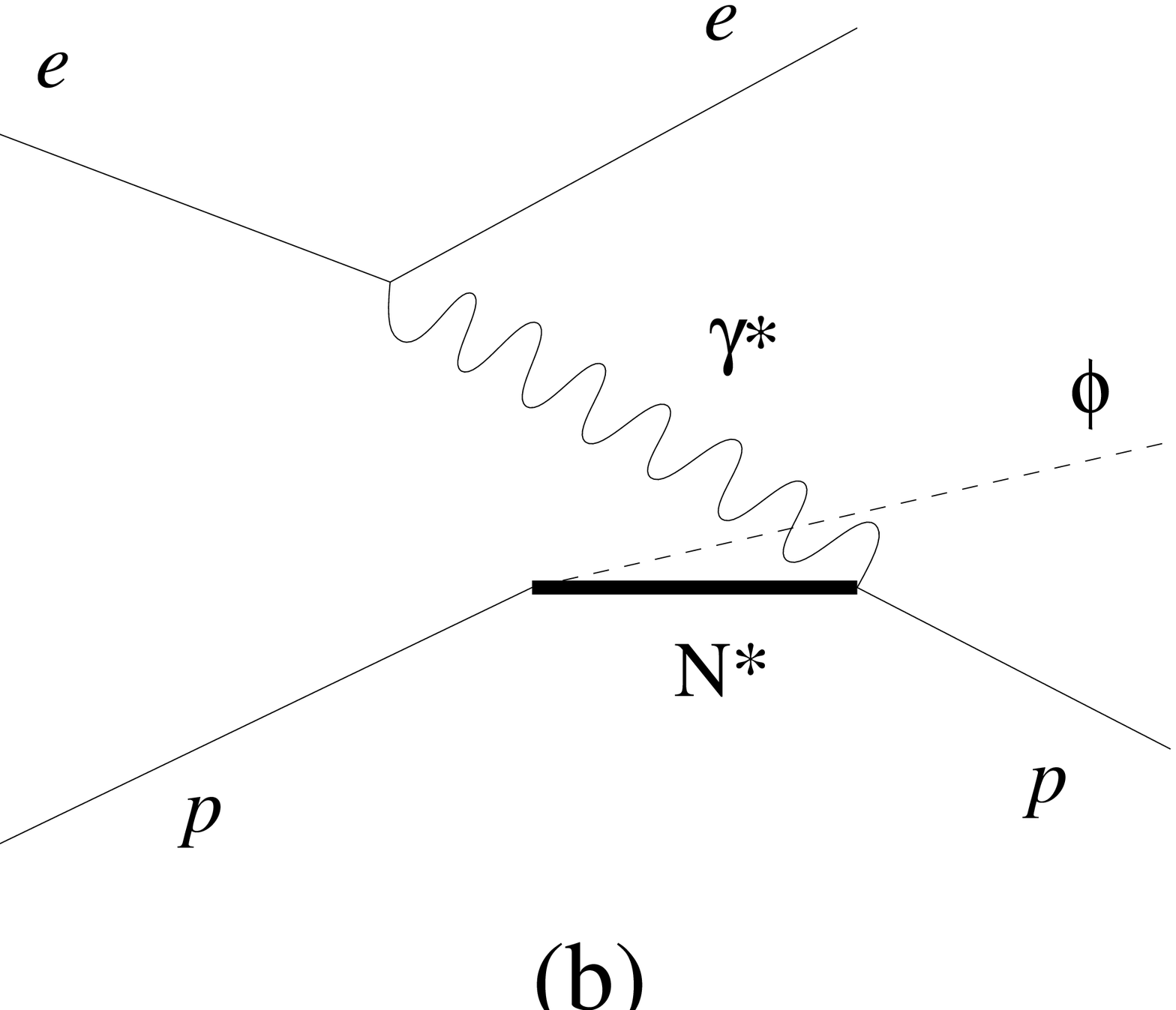}}
\vskip 1.2cm
\caption{$\phi$ meson production with some intermediate
     hadronic exited states.}
\label{pros2}
\end{figure}

Figure \ref{pros1}(b) corresponds to the process where an $s \bar s$
pair is directly knocked out by the photon and Fig. \ref{pros1}(c) to
the direct $uud$-knockout. It is also possible that the system would
have some hadronic intermediate states like $N$ and $N^\ast$ before
and after the $\phi$ meson is emitted as shown in Fig. \ref{pros2}.
These diagrams represent some of the rescattering effects and deserves
to be studied. However, we will focus only on {\em the direct knockout
mechanism} of Fig. \ref{pros1}(b,c) in this paper and leave the other
for future study. Instead of triple and double differential
electroproduction cross sections $\d^3 \sigma/ \d W \d Q^2 \d t$ and
$\d^2 \sigma/ \d W \d Q^2$ for the knockout process, respectively, we
will work with the cross sections $\sigma (Q^2,W)$ and
$\sigma (Q^2,W,t)$ defined as
\begin{eqnarray}
\sigma (Q^2,W,t) &=& \frac{1}{2\pi\Gamma_W(Q^2,W)}
\frac{\d^3 \sigma}{\d W \d Q^2 \d t}, \nonumber\\
\sigma (Q^2,W) &=& \frac{1}{2\pi\Gamma_W(Q^2,W)}
\frac{\d^2 \sigma}{\d W \d Q^2}.
\end{eqnarray}

The knockout amplitude $T_{fi}$ in the one-photon exchange
approximation may be written in the most general form as
\begin{eqnarray}
- \I (2\pi)^4 \delta^4 (p+q-p'-q_\phi) T_{fi}= \left\langle h_f\vert
\hat J^{h}_{\mu} \vert h_{i} \right\rangle \frac{g^{\mu \nu}}{q^2}
\left\langle k' \vert \hat J^{e}_{\nu} \vert k \right\rangle,
\end{eqnarray}
where $\left\langle f \vert \hat J^{h,e}_{\nu} \vert i \right\rangle$
are the hadron and electron electromagnetic (e.m.) current matrix
elements, respectively. The electron matrix element is given by
\begin{eqnarray}
\left\langle k' \vert \hat J^{e}_{\nu} \vert k \right\rangle \equiv
\sqrt{ \frac{M_e^2}{E_e E_e'} } \,
j^{e}_{\nu} = \sqrt{\frac{M_e^2}{E_e E_e'}} \,
\overline{u}_{m_{e'}} (k') e \gamma_{\nu} u_{m_e} (k),
\label{em:eltn}
\end{eqnarray}
where $u_{m_e}(k)$ is the plane wave electron Dirac spinor ($m_e$
denotes the spin projection) normalized as $\overline{u}u=1$. The
hadron e.m. current matrix element depends on the model for
description of the initial and final hadron states $\vert h_{i,f}
\rangle$ and the form of the e.m. current operator
$\hat J^{h}_{\mu}$. The additivity of the e.m. current in the
quark model enables one to write the amplitude $T_{fi}$ as
\begin{eqnarray}
T_{fi}=T_{fi}^{s\bar s}+T_{fi}^{uud},
\end{eqnarray}
where the first term describes the interaction of the e.m.
field with the $s$ and $\bar s$ quarks, i.e., the $s\bar s$-cluster
knockout, while the second one corresponds to the $uud$-knockout.

Then the squared amplitude consists of three terms which are the
contributions of the $s\bar s$- and $uud$-cluster knockout and the
interference:
\begin{equation}
\overline {\overline {\vert T_{fi}\vert^2}}
= \overline {\overline {\vert T_{fi}^{s\bar s}\vert^2}} + \overline
{\overline {\vert T_{fi}^{uud}\vert^2}} + \overline {\overline {\vert
T_{fi}^{\rm int}\vert^2}} .
\end{equation}


\section{Proton wavefunction}

For simplicity and for our qualitative study, we approximate the proton
wavefunction (\ref{Fockp}) as
\begin{equation}
|p\rangle = A | uud \rangle + B | uuds\bar s \rangle,
\label{prwf}
\end{equation}
by absorbing the contributions from $A_X$ terms into the coefficient
$A$ and keeping the leading order term in $B_X$ types. The $A_X$-type
terms {\em must} be present in the proton wavefunction. However, as
we shall see, the cross sections of the knockout mechanism for $\phi$
electroproduction depends on $(AB)^2$, so that such an approximation
may be justifiable in this process. We can further decompose the
wavefunction as
\begin{equation}
|p\rangle = A | [uud]^{1/2} \rangle + B \left\{ a_0 | \bbox{[} [
uud]^{1/2}
\otimes [s\bar s]^0 \bbox{]}^{1/2} \rangle + a_1 | \bbox{[} [ uud]^{1/2}
\otimes [s\bar s]^1 \bbox{]}^{1/2} \rangle \right\},
\label{protonwf}
\end{equation}
where the superscripts denote the spin of each cluster. Then $B^2$
is the strangeness admixture of the proton and $a_0^2$ and $a_1^2$
correspond to the spin-0 and spin-1 fractions of $s\bar s$ cluster,
respectively. They are constrained to $A^2+B^2=a_0^2+a_1^2=1$ by the
normalization of the wavefunction. The symbol $\otimes$ represents
a possible orbital angular momentum between the two clusters. In the
simplest picture, the $s$ and $\bar s$ quarks are in a relative $1s$
state with respect to each other with negative intrinsic parity. The
$uud$ cluster is also in a relative $1s$ state with respective to the
CM of the cluster. We also neglect a possible hidden color components,
which was shown to be negligible in SU(2) 5-quark model \cite{FH}.
Therefore, the configuration (\ref{protonwf}) corresponds to $\eta_s$
and $\phi$ meson ``in the air" in proton wavefunction, where $\eta_s$
(=$s\bar s$) is the mixture of $\eta$ and $\eta'$.

Then to describe positive parity proton, the $s\bar s$ cluster should
be in a relative $P$-wave about the CM with the $uud$ cluster that is
the ``bare" proton. More complicated configurations are possible by
allowing complex combinations. But we expect that the above two
components give major contribution to the $\phi$ electroproduction.
Also excluded is the higher spin states of $uud$ cluster. For example,
we may include spin 3/2 $uud$ cluster. However, since the isospin of
$\phi$ is zero, the $uud$ cluster should have isospin 1/2.
Experimentally observed states with $i=1/2$, $j=3/2$ are $N^*(\frac32^-)$
and $N^*(\frac32^+)$ at 1520 MeV and 1720 MeV, respectively. Because of
their heavier mass, their role is expected to be small and excluded in
our study as in Ref. \cite{HKW}.

The spin-orbital wavefunction can be obtained by noting that the total
spin $\bbox{J}$ (with $j=1/2$) is
\begin{equation}
\bbox{J} = \bbox{J}_c + \bbox{J}_{s \bar s} = \bbox{J}_{uud} + \bbox{L}
+ \bbox{J}_{s\bar s},
\end{equation}
where $\bbox{J}_{uud}$ is the spin of the $uud$ cluster $(j_{uud}=1/2)$,
$\bbox{J}_{s\bar s}$ of the $s\bar s$ cluster ($j_{s\bar s}=$0,1), and
$\bbox{L}$ the relative orbital angular momentum $(\ell=1)$.
Therefore, we can write the spin wavefunction as
\begin{equation}
\textstyle
| j=\frac12, m_i \rangle = {\displaystyle \sum_{\mbox{\scriptsize $m$'s}}}
\langle \frac12\, m' \,\,
1\, m_\lambda | j_c\, m_c \rangle \langle j_c\, m_c \,\, j_{s\bar s} \,
m_{s\bar s} | \frac12\, m_i \rangle
| \frac12\, m' \rangle_{uud} | j_{s\bar s} \, m_{s\bar s} \rangle_{s\bar s}
| 1 \, m_\lambda \rangle_\ell.
\end{equation}
For $j_{s \bar s} =0$, we have $j_c=\frac12$ and
\begin{equation}
\textstyle
| \frac12, m_i ; j_{s \bar s} =0 \rangle =
{\displaystyle \sum_{\mbox{\scriptsize $m$'s}} } \langle
\frac12\, m' \,\,
1\, m_\lambda | \frac12\, m_i \rangle | \frac12\, m_i \rangle_{uud} | 0\,
0\rangle_{s\bar s} | 1 \, m_\lambda \rangle_\ell.
\end{equation}
And for  $j_{s \bar s} =1$, since $j_c$ can be either 1/2 or 3/2, we have
\begin{eqnarray}
{\textstyle
| \frac12, m_i ; j_{s \bar s} =1 \rangle} &=&
\sum_{j_c} b_{2j_c} \sum_{\scriptsize \mbox{$m$'s}}
\textstyle
\langle \frac12\, m' \,\, 1\, m_\lambda | j_c \, m_c \rangle
\langle j_c \, m_c \,\, 1\, m_{s\bar s} |\frac12\, m_i \rangle
| \frac12\, m' \rangle_{uud} | 1 \, m_{s\bar s} \rangle_{s\bar s}
| 1 \, m_\lambda \rangle_\ell,
\nonumber \\
\end{eqnarray}
with $b_1^2+b_3^2=1$. Finally, the spin-orbital wavefunction reads
\begin{equation}
\textstyle
| \frac12, m_i \rangle = A | \frac12, m_i \rangle\!\rangle + B \{ a_0 |
\frac12, m_i; j_{s\bar s}=0 \rangle + a_1 | \frac12, m_i; j_{s\bar s}=1
\rangle \},
\label{proton-wf}
\end{equation}
where $| \frac12, m_i \rangle\!\rangle$ denotes the bare proton
wavefunction.  When combined with the flavor, spatial and color
wavefunctions, this completes our proton wavefunction.


\section{Non-relativistic quark model}

\subsection{The model}

For the spatial wavefunctions of hadrons, we use the nonrelativistic
harmonic oscillator quark potential model in this Section. If we consider
the proton wavefunction in 5 quark configurations, the spatial
wavefunction is obtained from the Hamiltonian
\begin{equation}
H^{(5)} = \sum_{i=1}^5 \left( - \frac{\bbox{\nabla}_i^2}{2M_i} \right)
+ k \sum_{i\neq j=1}^3 ( \bbox{x}_i - \bbox{x}_j )^2
+ k^s  ( \bbox{x}_4 - \bbox{x}_5 )^2
+ k' \sum_{{i=1,2,3}\atop{j=4,5}} ( \bbox{x}_i - \bbox{x}_j)^2,
\end{equation}
where the labels $i=1,2,3$ refer to the particles in the $uud$-cluster and
$i=4,5$ to the $s\bar s$-cluster, and $M_i$ is the mass of $i$-th quark. In
this work, we use $M_q \equiv M_u = M_d = 330$ MeV and $M_s = 500$ MeV.
This Hamiltonian can be diagonalized by introducing Jacobian coordinates as
\begin{eqnarray}
&& \bbox{\xi}_1 = \frac{1}{\sqrt6} ( \bbox{x}_2 + \bbox{x}_3 - 2
\bbox{x}_1 ), \qquad
\bbox{\xi}_2 = \sqrt{\frac{1}{2}} ( \bbox{x}_3 - \bbox{x}_2 ),
\qquad
\bbox{\rho} = \sqrt3 (\bbox{x}_5 - \bbox{x}_4), \nonumber \\
&& \bbox{\chi} = \frac13 ( \bbox{x}_1 + \bbox{x}_2 + \bbox{x}_3 ) -
\frac12 ( \bbox{x}_4 + \bbox{x}_5), \qquad
\bbox{R} = ( \sum_i M_i \bbox{x}_i ) / M_5,
\label{Jac}
\end{eqnarray}
where $M_5 = \sum_i M_i = 3 M_q + 2 M_s$.

Then the spatial wavefunction  $\Psi^{\text{SP}}_{\bbox{P},\lambda}
(\bbox{r}_1,\bbox{r}_2,\bbox{r}_3,\bbox{r}_4, \bbox{r}_5)$ with
momentum $\bbox{P}$ and the projection $\lambda$  of the orbital
angular momentum $\ell$(=1), has the form
\begin{mathletters} \label{SHO-wf}
\begin{equation}
\Psi^{\rm SP}_{\bbox{P}, \lambda}(\bbox{r}_1,\bbox{r}_2,
\bbox{r}_3,\bbox{r}_4,\bbox{r}_5) =
\e^{\I \bbox{P} \cdot \bbox{R}}
\psi^{uud} (\bbox{\xi}_1,\bbox{\xi}_2) \psi^{s\bar s} (\bbox{\rho})
\psi_\lambda(\bbox{\chi}) =
\e^{\I \bbox{P} \cdot \bbox{R}} 
\psi(\bbox{\xi}_1) \psi(\bbox{\xi}_2) \psi(\bbox{\rho})
\psi_\lambda(\bbox{\chi}),
\end{equation}
where the normalized radial wavefunctions are
\begin{eqnarray}
\psi (\bbox{u}_i) &=& \left( \frac{\Omega_i}{\pi} \right)^{3/4}
\e^{-\frac12
\Omega_i \bbox{u}_i^2}, \nonumber \\
\psi_\lambda(\bbox{\chi}) &=& \sqrt{2\Omega_\chi} \left(
\frac{\Omega_\chi}{\pi} \right)^{3/4} \chi_\lambda \e^{-\frac12 \Omega_\chi
\bbox{\chi}^2},
\end{eqnarray}
where $\bbox{u}_i=\bbox{\xi}_1,\bbox{\xi}_2,\bbox{\rho}$ and $\Omega_i$
represents $\Omega_{u_i}$. They are normalized as $\int d^3 \bbox{u}_i \, |
\psi(\bbox{u}_i)|^2 = \int d^3 \bbox{\chi} |\psi_\lambda
(\bbox{\chi})|^2 =1$, and
$\chi_0 = \chi_z$, $\chi_{\pm 1} = \mp ( \chi_x \pm \I \chi_y )/\sqrt2$.
The dimensional parameters are
\begin{equation}
\Omega^2_{\xi_1} = \Omega^2_{\xi_2} \equiv \Omega^2_{\xi} = 2 ( 3 k
+ 2 k') M_q , \qquad
\Omega_\rho^2 = (k^s + {\textstyle\frac32} k') M_s, \qquad
\Omega_\chi = 12 M_\chi k',
\end{equation}
with $M_\chi = (6M_q M_s)/M_5$.
The spatial wavefunctions of the bare proton and
$\phi$ meson have the similar structure.

\end{mathletters}

We fix the dimensional parameters as in Ref. \cite{HKW} by using the fact
that they are related to the hadron rms radii. The wavefunctions
(\ref{SHO-wf}) give
\begin{equation}
\langle r_{uud}^2 \rangle = \frac{1}{\Omega_\xi}, \qquad
\langle r_{s\bar s}^2 \rangle = \frac38 \frac{1}{\Omega_\rho}, \qquad
\langle \chi^2 \rangle = \frac52 \frac{1}{\Omega_\chi},
\end{equation}
and we assume that $\Omega_\xi^0$ of the bare proton and $\Omega_\rho^0$ of
the $\phi$ are equal to $\Omega_\xi$ and $\Omega_\rho$, respectively.
By making use of the empirical values of $\sqrt{\langle r_p^2 \rangle}$
($=0.83$ fm) and $\sqrt{\langle r_\phi^2 \rangle}$ ($=0.45$ fm) and by
introducing scaling factor $c$ (=1.5) \cite{HKW}, we get
\begin{equation}
\sqrt{\Omega_\xi} = 1.81 \mbox{ fm}^{-1}, \qquad
\sqrt{\Omega_\rho} = 2.04 \mbox{ fm}^{-1}, \qquad
\sqrt{\Omega_\chi} = 2.43 \mbox{ fm}^{-1}.
\end{equation}
To obtain $\Omega_\chi$ we assumed that the spring constant of $\chi$
coordinate is the same as that of $\phi$ meson.

We also use the additive form of the NRQM e.m. current;
\begin{equation}
\hat J_0^h = \sum_{k=1}^5 \hat J_{k,0}^h, \hskip 1cm
\hat{\!\!\bbox{J}}^h = \sum_{k=1}^5 \,\,\hat{\!\!\bbox{J}}_k^h,
\end{equation}
where
\begin{eqnarray}
\langle p_k' | \hat J_{k,0}^h | p_k \rangle &=& e_k \delta^3 ( \bbox{p}'_k
- \bbox{p}_k) \nonumber \\
\langle p_k' | \,\,\hat{\!\!\bbox{J}}_k^h | p_k \rangle &=& \frac{e_k}{2M_k}
\left\{ \bbox{p}_k' + \bbox{p_k} + i \bbox{\sigma}_k \times (
\bbox{p}_k' - \bbox{p}_k ) \right\},
\label{EM:NRQM}
\end{eqnarray}
in momentum space, where $e_k$ and $M_k$ are the charge and mass of the
$k$-th quark and $\bbox{p}_k'$ ($\bbox{p}_k$) is its final (initial)
momentum.

\subsection{$s \bar s$-knockout}

The $s \bar s$-knockout process is depicted in Fig.~\ref{pros1}(b).
Because of the symmetric property of the spatial wavefunction and
the current, it is manifest that {\em only the $j_{s \bar s}=0$ part
of the proton wavefunction (\ref{protonwf}) and the magnetic part
of the e.m. current (\ref{EM:NRQM}) can contribute.\/} Then the
$T$-matrix is obtained as
\begin{equation}
T_{s \bar s}^{\rm (NR)} = - {\textstyle \frac23}
A^* B a_0 \left( \frac{e\mu_s}{2M_N} \right)
\sum_\lambda I_\lambda^{s\bar s}
{\textstyle\langle \frac12 \, m_f \, 1 \, \lambda | \frac12 \, m_i \rangle}
(-1)^{m_\phi} ( \bbox{q} \times \bbox{\cal A} )_{-m_\phi},
\label{Tnrss}
\end{equation}
where $\mu_s=M_N/M_s$ and ${\cal A}_\mu$ is a four-vector defined as
\begin{eqnarray}
{\cal A}_\mu = \frac{1}{Q^2} \sqrt{\frac{M_e^2}{E_e E_e'}} \,
\overline{u}_{m_{e'}} (k') e \gamma_{\mu} u_{m_e} (k).
\end{eqnarray}
The spatial overlap integral $I_\lambda^{s\bar s}$ in NRQM is defined as
\begin{mathletters}
\begin{equation}
I_\lambda^{s \bar s} = \gamma_{s \bar s}
\beta_{s \bar s} (\bbox{q})
\tilde \psi^{\rm (NR)}_\lambda (\bbox{p}'),
\end{equation}
where
\begin{eqnarray}
\gamma_{s \bar s} &=& \int \d \bbox{\xi}_1 \d \bbox{\xi}_2 \,
\psi^{uud}_f (\bbox{\xi}_1, \bbox{\xi}_2 )
\psi^{uud}_i (\bbox{\xi}_1, \bbox{\xi}_2 ),
\\
\beta_{s \bar s} (\bbox{q}) &=& \int \d \bbox{\rho}\,
\e^{\I \bbox{q} \cdot \bbox{\rho}/2}
\psi_f^{s \bar s} (\bbox{\rho}) \psi_i^{s \bar s} (\bbox{\rho}),
\\
\tilde\psi^{\rm (NR)}_\lambda(\bbox{p}') &=& \int \d \bbox{\chi} \,
\e^{-\I \bbox{p}' \cdot \bbox{\chi} }
\psi_\lambda (\bbox{\chi}).
\end{eqnarray}

By making use of
\end{mathletters}
\begin{equation}
\sum\limits_{m_e,m_{e'}}
{\cal A}_\mu{\cal A}_\nu^{\ast} = \frac{e^2}{q^4} \frac{1}{E_eE_e'}
[k_\mu k_\nu'+ k_\mu' k_\nu - g_{\mu \nu} (k \cdot k'-M_e^2)],
\end{equation}
for unpolarized case, we can obtain
\begin{equation}
\overline{\overline{|T_{s \bar s}^{\rm (NR)}|^2}} =
\frac49
(ABa_0)^2 \frac{e^4}{Q^4}
\Gamma_{s \bar s} F_{s \bar s} (\bbox{q})
V_{s \bar s} (\bbox{p}')
\cos^2 {\textstyle \frac{\theta}{2}}
\left( f_1^{s \bar s} + f_2^{s \bar s}
\tan^2 {\textstyle \frac{\theta}{2}} \right),
\end{equation}
where
\begin{equation}
f^{s \bar s}_1 = \mu_s^2\eta, \qquad
f^{s \bar s}_2 = 2 \mu_s^2[ \eta + ( \eta + z )^2 ].
\end{equation}
The spatial integrals are calculated using (\ref{SHO-wf}) as
\begin{eqnarray}
\Gamma_{s \bar s} &=& \gamma_{s \bar s}^2 = 1,
\nonumber \\
F_{s \bar s} (\bbox{q}) &=& [ \beta_{s\bar s} (\bbox{q}) ]^2
= \exp\left[-\bbox{q}^2/(8\Omega_\rho)\right],
\nonumber \\
V_{s \bar s} (\bbox{p}') &=& \frac13
\sum_{\lambda} | \tilde \psi_\lambda
(\bbox{p}') |^2 = \frac{2(2\pi)^3}{3} \left( \frac{1}{\pi \Omega_\chi}
\right)^{3/2} \frac{\bbox{p}^{\prime 2}}{\Omega_\chi} \,
\e^{-\bbox{p}^{\prime 2} / \Omega_\chi}.
\label{GFV}
\end{eqnarray}
This is the result derived in Ref. \cite{HKW} for the $s\bar s$ knockout
process. Since it depends on $t$ only through $\bbox{p}^{\prime 2}$ of
$V_{s \bar s}$, one can see that the cross section is maximum near
$t_{\rm max}$ by noting that $\bbox{p}^{\prime 2}$ increases as
decreasing $t$.

\subsection{$uud$-knockout}

As depicted in Fig. \ref{pros1}(c), the $s\bar s$ cluster is a spectator
in this process. So {\em only the $j_{s \bar s}=1$ part
contribute\/} contrary to the $s \bar s$ knockout. The relevant
amplitude reads
\begin{equation}
T_{uud}^{\rm (NR)} = \I A^* B a_1 \sum_{j_c,m',m_c,\lambda} b_{2j_c}
{\textstyle
\langle \frac12 \, m' \, 1 \, \lambda | j_c \, m_c \rangle
\langle j_c \, m_c \, 1 \, m_\phi | \frac12 \, m_i \rangle
\langle \frac12 \, m_f | F_\mu^{uud} | \frac12 \, m' \rangle}
I_\lambda^{uud} {\cal A}^\mu,
\label{Tnruud}
\end{equation}
where
\begin{equation}
F^{uud}_0 = e, \qquad
\bbox{F}^{uud} = \displaystyle \frac{e}{2M_N} (2 \bbox{p}' - \bbox{q}
+ \I \mu \bbox{\sigma} \times \bbox{q} ),
\label{FF}
\end{equation}
with $\mu = M_N/M_q$.
The overlap integral $I_\lambda^{uud}$ reads
\begin{mathletters}
\begin{equation}
I_\lambda^{uud} = \gamma_{uud} \beta_{uud} (\bbox{q})
\tilde \psi^{\rm (NR)}_\lambda (-\bbox{q}_\phi),
\end{equation}
where
\begin{eqnarray}
\gamma_{uud} &=& \int \d \bbox{\rho} \, \psi_f^{s \bar s} (\bbox{\rho})
\psi_i^{s \bar s} (\bbox{\rho}),
\\
\beta_{uud} (\bbox{q}) &=& \int \d \bbox{\xi}_1 \d \bbox{\xi}_2 \,
\e^{-\I \sqrt{\frac23} \bbox{q} \cdot \bbox{\xi}_1}
\psi_f^{uud} (\bbox{\xi}_1, \bbox{\xi}_2) \psi_i^{uud} (\bbox{\xi}_1,
\bbox{\xi}_2),
\\
\tilde\psi^{\rm (NR)}_\lambda (-\bbox{q}_\phi) &=&
\int \d \bbox{\chi} \, \e^{\I \bbox{q}_\phi \cdot
\bbox{\chi}} \psi_\lambda (\bbox{\chi}).
\end{eqnarray}

Then we have the squared amplitude as
\end{mathletters}
\begin{equation}
\overline{\overline{|T^{\rm (NR)}_{uud}|^2 }} 
= ( ABa_1)^2 \frac{e^4}{Q^4}
\Gamma_{uud} F_{uud}(\bbox{q}) V_{uud} (\bbox{q}_\phi)
\cos^2 {\textstyle \frac{\theta}{2}}
\left( f_1^{uud} + f_2^{uud} \sin^2 {\textstyle \frac{\theta}{2}}
\right),
\end{equation}
where
\begin{eqnarray}
f_1^{uud} &=& \eta \mu^2 \left\{ 1 + \frac{1}{\eta \mu^2} \left( 1 -
\frac{\nu\tilde \nu}{\bbox{q}^2} \right)^2 + \frac{c}{\mu^2} \right\},
\nonumber \\
f_2^{uud} &=& 2 (\mu^2 +c ) [ \eta + (z+\eta)^2 ],
\label{f12uud}
\end{eqnarray}
with
\begin{equation}
\tilde\nu = \frac{1}{2M_N} ( 2 \bbox{p}' \cdot \bbox{q} - \bbox{q}^2 ),
\qquad
c = -\frac{1}{2\bbox{q}^4} \lambda( \bbox{q}^2, \bbox{q}_\phi^2,
\bbox{p}^{\prime 2} ),
\label{def:c}
\end{equation}
where $\lambda(x,y,z)$ is defined in Eq. (\ref{lambda}).
The radial wavefunctions of NRQM give
\begin{eqnarray}
\Gamma_{uud} &=& \gamma^2_{uud} = 1,
\nonumber \\
F_{uud} (\bbox{q}) &=& [ \beta_{uud} (\bbox{q}) ]^2
= \exp \left[ - \bbox{q}^2/(3\Omega_\xi) \right],
\nonumber \\
V_{uud} (\bbox{q}_\phi) &=& \frac13 \sum_\lambda | \tilde \psi_\lambda
(-\bbox{q}_\phi) |^2 = \frac{2(2\pi)^3}{3} \left( \frac{1}{\pi \Omega_\chi}
\right)^{3/2} \frac{\bbox{q}_\phi^2}{\Omega_\chi} \,
\e^{-\bbox{q}_\phi^2 / \Omega_\chi}.
\label{FVnrud}
\end{eqnarray}
Then the cross section depends on $t$ only through $\bbox{q}_\phi^2$ of
$V_{uud}$.  This shows that in contrary to the $s \bar s$-knockout, the
$uud$-knockout process gives its main contribution near $t_{\rm min}$.

\subsection{gauge invariance}

The electron e.m. current in (\ref{em:eltn}) satisfies the
conservation condition $q^\mu j^{e}_{\mu} = 0$. The gauge invariance
implies the same condition for the hadronic current:
\begin{eqnarray}
q^\mu\hat J^{h}_{\mu} = 0.
\label{gauge}
\end{eqnarray}
In general, however, this condition is not satisfied in inelastic
scattering \cite{Akerlof}. In our case, this condition is satisfied only
in the $s \bar s$ knockout process since only the magnetic part of the e.m.
current contributes. In $uud$ knockout process, however, the convection
current takes part in the process and the relation (\ref{gauge}) is not
obeyed. The breakdown of the gauge invariance is crucial at small $Q^2$
because the matrix elements of $q_\mu J^{h}_{\mu}$
are proportional to $(\nu - \tilde \nu)^2/Q^2$ as $Q^2 \to 0$.

The most commonly used technique of enforcing gauge invariance
\cite{FKN,Akerlof} in photoproduction is to project out the gauge
non-invariant part as
\begin{equation}
{F}_\mu \to
{F}_{\mu}'=(g_{\mu\nu}-\frac{q_\mu q_\nu}{q^2}) {F}^{\nu}.
\end{equation}
This modification, however, is not adequate for the electroproduction,
where the electron e.m. current cancels the subtracted part.

A possible way for restoring the gauge invariance is to modify the
longitudinal component of the current. Decomposition of the spatial
component of the current (\ref{FF}) gives the longitudinal part as
\begin{equation}
\bbox{F}_\parallel = \frac{e \tilde \nu}{\bbox{q}^2} \bbox{q},
\end{equation}
and we modify it as
\begin{equation}
\bbox{F}_\parallel \to \bbox{F}'_\parallel = \frac{e\nu}{\bbox{q}^2}
\bbox{q}.
\label{modif}
\end{equation}
This {\rm ansatz\/} restores the gauge invariance of the hadronic e.m.
current.

The modification (\ref{modif}) causes a change of $f_1^{uud}$ in the
squared $T$ matrix as
\begin{equation}
f_1^{uud} \to \tilde f_1^{uud} = \eta \mu^2 \left( 1 + \frac{1}{\mu^2}
\frac{\eta}{[\eta + (\eta+z)^2 ]^2} + \frac{c}{\mu^2}, \right),
\end{equation}
with others unchanged. This is equivalent to replace $\tilde \nu$ by $\nu$
in Eq. (\ref{f12uud}). Close inspection shows that the second and
the third terms are negligible in the considered kinematical region. This
leads to the conclusion that {\em the main contribution to the $uud$
knockout process comes also from the magnetic part of the e.m. current.\/}%
\footnote{
One may modify the time component of the current to restore the gauge
invariance. But it does not change this conclusion.}

\subsection{interference}

{}From the $T$ matrices of $s \bar s$ knockout (\ref{Tnrss}) and of $uud$
knockout (\ref{Tnruud}), we can obtain the interference between the
two processes;
\begin{equation}
\overline{\overline{|T_{\rm int}^{\rm (NR)}|^2}} =
c_{\rm int} (AB)^2 a_0 a_1 \frac{e^4}{Q^4}
\Gamma_{\rm int} F_{\rm int} (\bbox{q})
V_{\rm int} (\bbox{p}' , \bbox{q}_\phi)
\cos^2 {\textstyle \frac{\theta}{2}}
\left( f_1^{\rm int} + f_2^{\rm int}
\tan^2 \textstyle{\frac{\theta}{2}} \right),
\end{equation}
where
\begin{eqnarray}
c_{\rm int} &=& \frac{4}{9\sqrt3} (b_1 + 2 \sqrt2 b_3 ),
\nonumber \\
f_1^{\rm int} &=& \mu\mu_s \eta,
\nonumber \\
f_2^{\rm int} &=& 2 \mu\mu_s [ \eta + (\eta+z)^2],
\end{eqnarray}
and
\begin{eqnarray}
\Gamma_{\rm int} &=& \gamma_{s \bar s} \gamma_{uud} = 1,
\nonumber \\
F_{\rm int} (\bbox{q} ) &=& \beta_{s \bar s} (\bbox{q}) \beta_{uud}
(\bbox{q}) = \exp( - \frac{8 \Omega_\rho + 3 \Omega_\chi}{48 \Omega_\rho
\Omega_\chi} \bbox{q}^2 ),
\nonumber \\
V_{\rm int} ( \bbox{p}' , \bbox{q}_\phi) &=& \frac13 \sum_\lambda \tilde
\psi^{\rm (NR)}_\lambda (\bbox{p}') [\psi^{\rm (NR)}_\lambda (- \bbox{q}_\phi
) ]^* = - \frac83 \left(
\frac{\pi}{\Omega_\chi} \right)^{3/2} \frac{\bbox{p}' \cdot \bbox{q}_\phi}
{\Omega_\chi} \e^{-( \bbox{p}^{\prime2} +
\bbox{q}_\phi^2 ) / 2 \Omega_\chi}.
\end{eqnarray}
Therefore, if we assume $c_{\rm int} > 0$, the interference between the two
knockout processes is constructive when $a_0a_1 <0$  and destructive when
$a_0 a_1 >0$.

\subsection{results}

For numerical calculations, we fix the incident electron energy $E_e$
as 11.5 GeV and $W=2.1$ GeV throughout this paper. We also use the
parameters of VDM and NRQM as determined before. The purpose of this
study is to determine the coefficients of the proton wavefunction
(\ref{protonwf}), especially $B^2$, by comparing the cross sections of
the knockout process with the VDM predictions. For our qualitative
study, we set $a_0^2=a_1^2 =1/2$ and $b_1=b_3=1/\sqrt2$. Since the
interference depends on the sign of $a_0 a_1$, we give only its magnitude.

\begin{figure}
\vskip 1cm
\centerline{\epsfxsize=0.48\hsize \epsffile{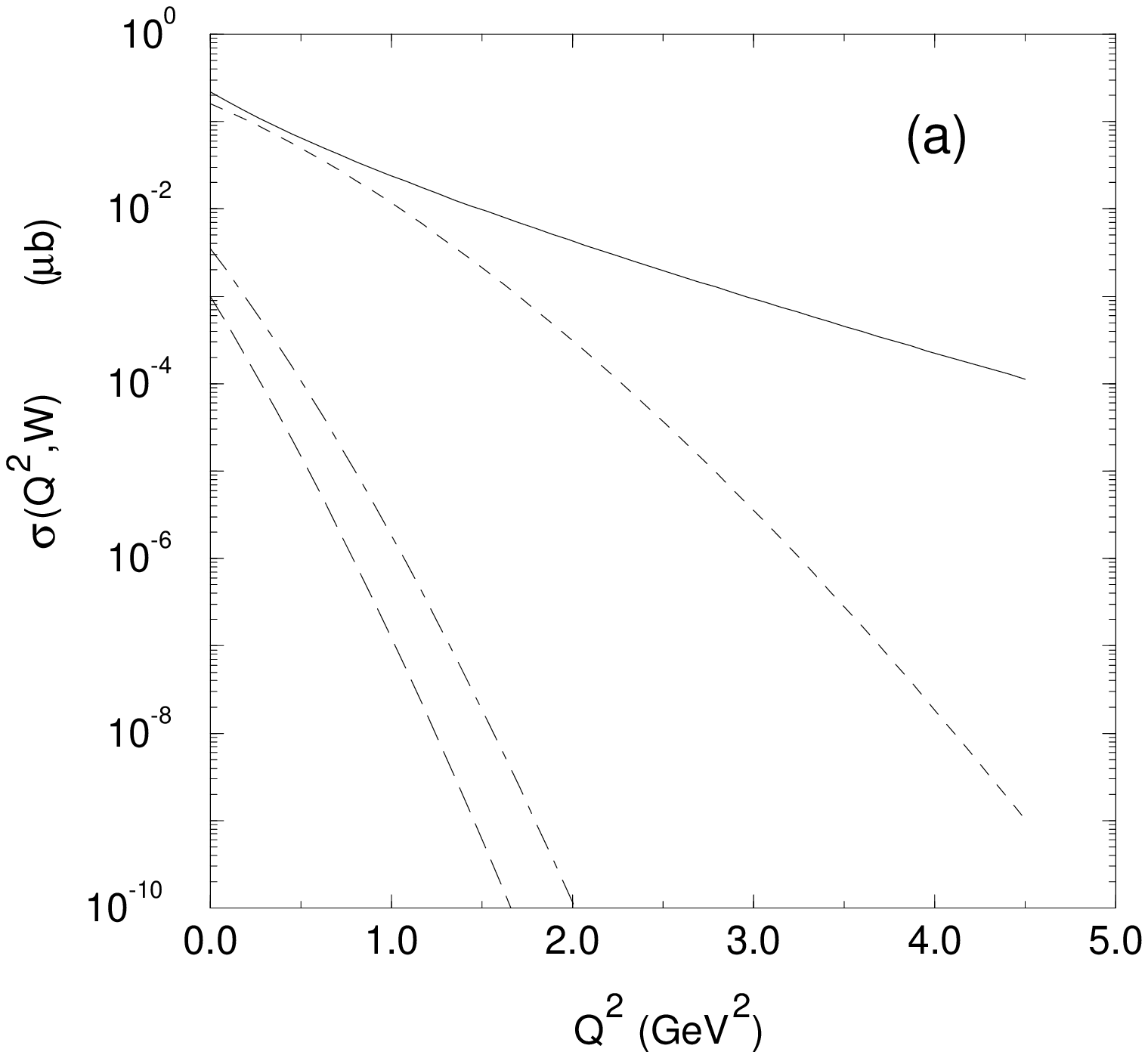}
 \hskip 2cm \epsfxsize=0.48\hsize \epsffile{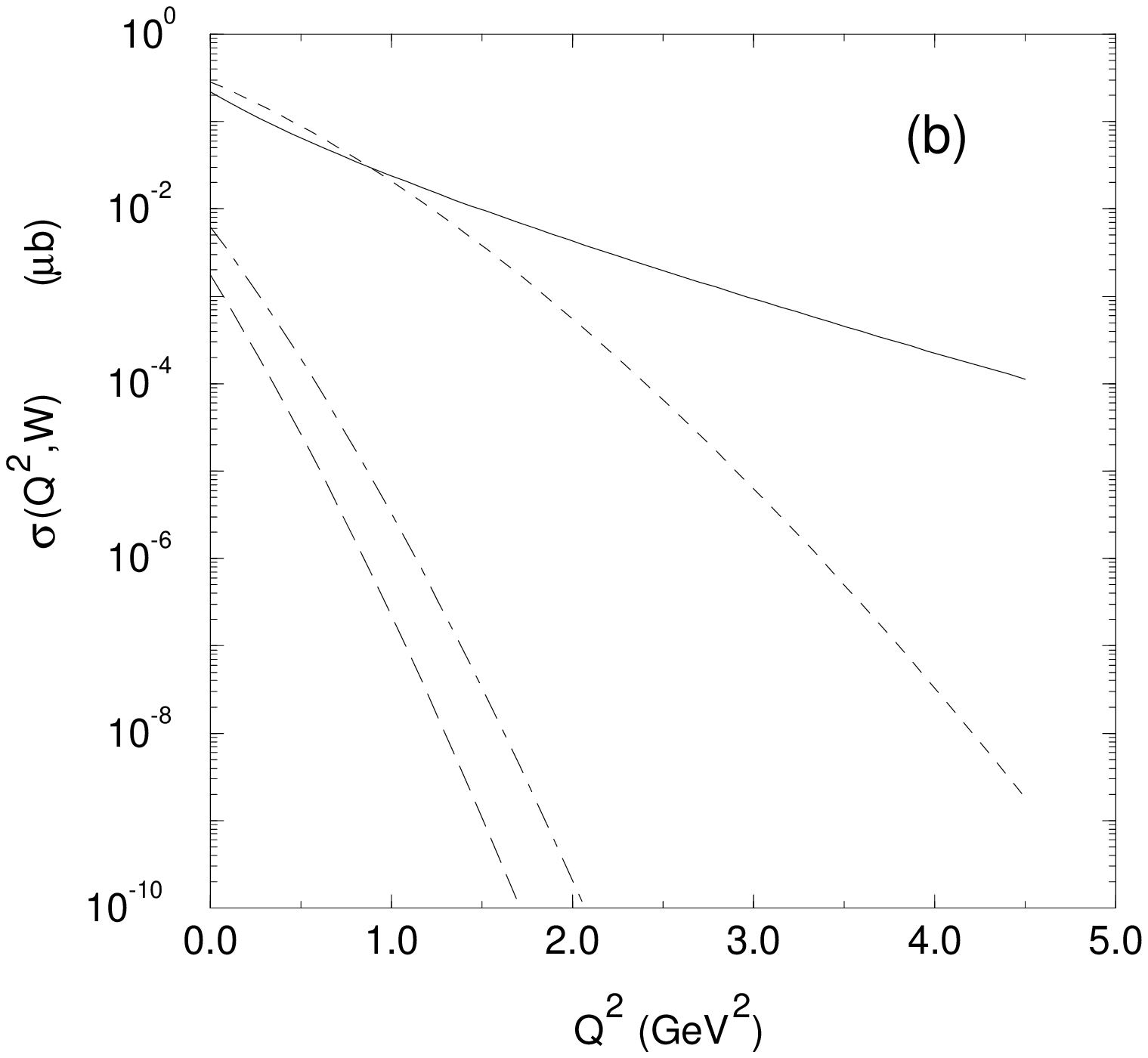}}
\caption{
The $Q^2$-dependence of the cross section
$\sigma(Q^2,W)$ at $W$=2.1 GeV and $E_e=11.5$ GeV in NRQM with (a)
$B^2=10$ \% and (b) $B^2=20$\%. The diffractive cross section is
given by the solid line, the $s \bar s$- and $uud$-knockout cross
sections are by the dotted and dashed lines, respectively, and the
interference term is by the dash-dotted line.}
\label{nrQ}
\end{figure}

Given in Fig.~\ref{nrQ} is $\sigma(Q^2,W)$ with (a) $B^2= 10$\%
and (b) $B^2=20$\%. The solid line is the VDM prediction and the dotted,
dashed, and dot-dashed lines are the contributions from $s \bar s$
knockout, $uud$ knockout, and the interference, respectively. This shows
that the knockout process is comparable to the VDM with $B^2=10$--20\% at
small $Q^2$, say, $Q^2 < 1.5$ GeV$^2$ \cite{HKW}. It is also manifest that
over the entire region of $Q^2$ the contribution of the $uud$ knockout and
the interference are suppressed with respect to that of the $s\bar s$
knockout process. This is because of the strong suppression of the form
factor $F_{uud}$ compared with $F_{s \bar s}$, where the former involves
$\langle r_{uud}^2 \rangle$ and the latter involves $\langle r_{s \bar s}^2
\rangle$. In general, the knockout cross section in NRQM falls rapidly
with $Q^2$ than that of VDM. This is mainly due to the strong
suppression of $F_{s \bar s, uud}$ of NRQM.

\begin{figure}
\vskip 1cm
\centerline{\epsfxsize=0.48\hsize \epsffile{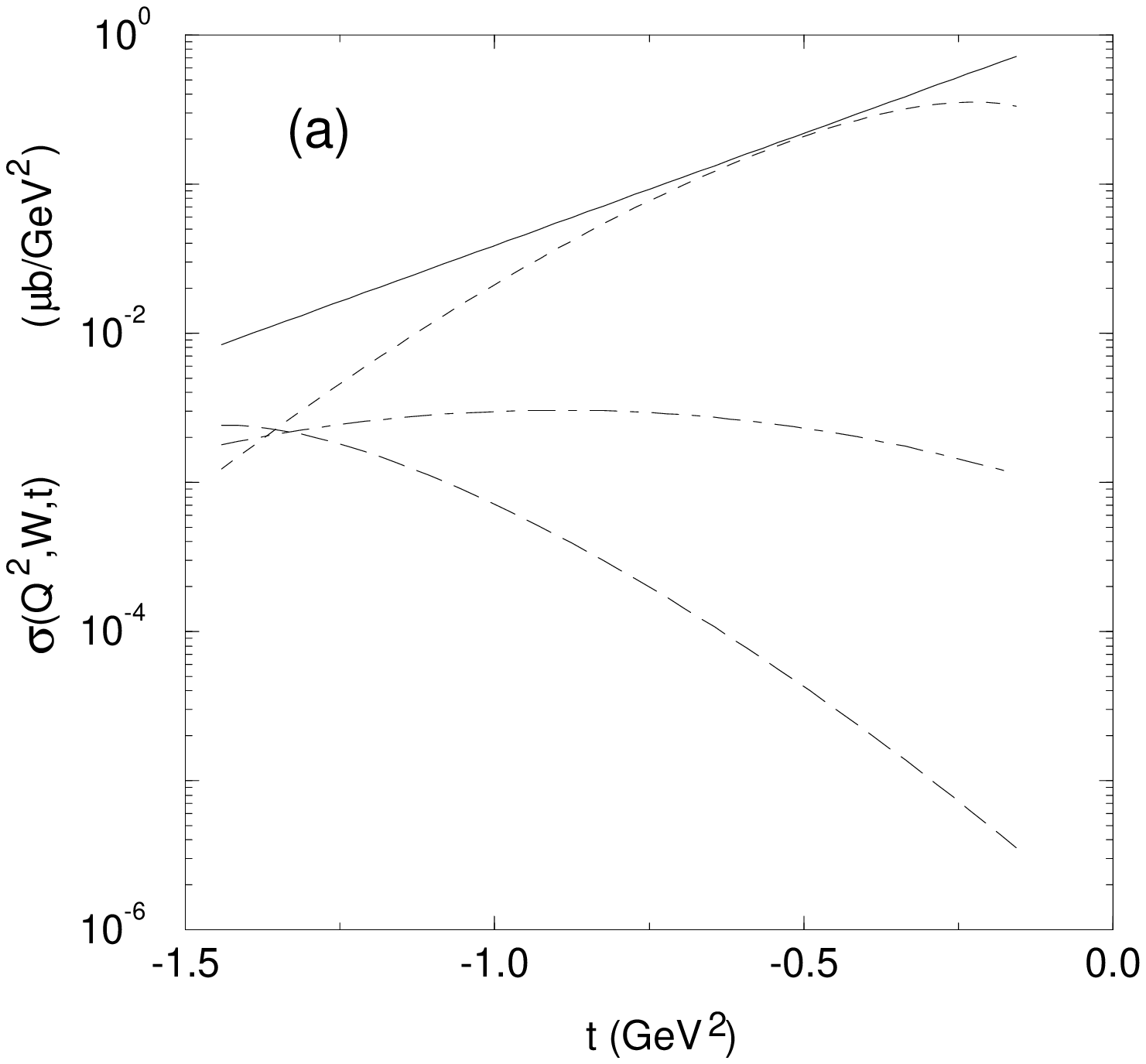}
 \hskip 2cm \epsfxsize=0.48\hsize \epsffile{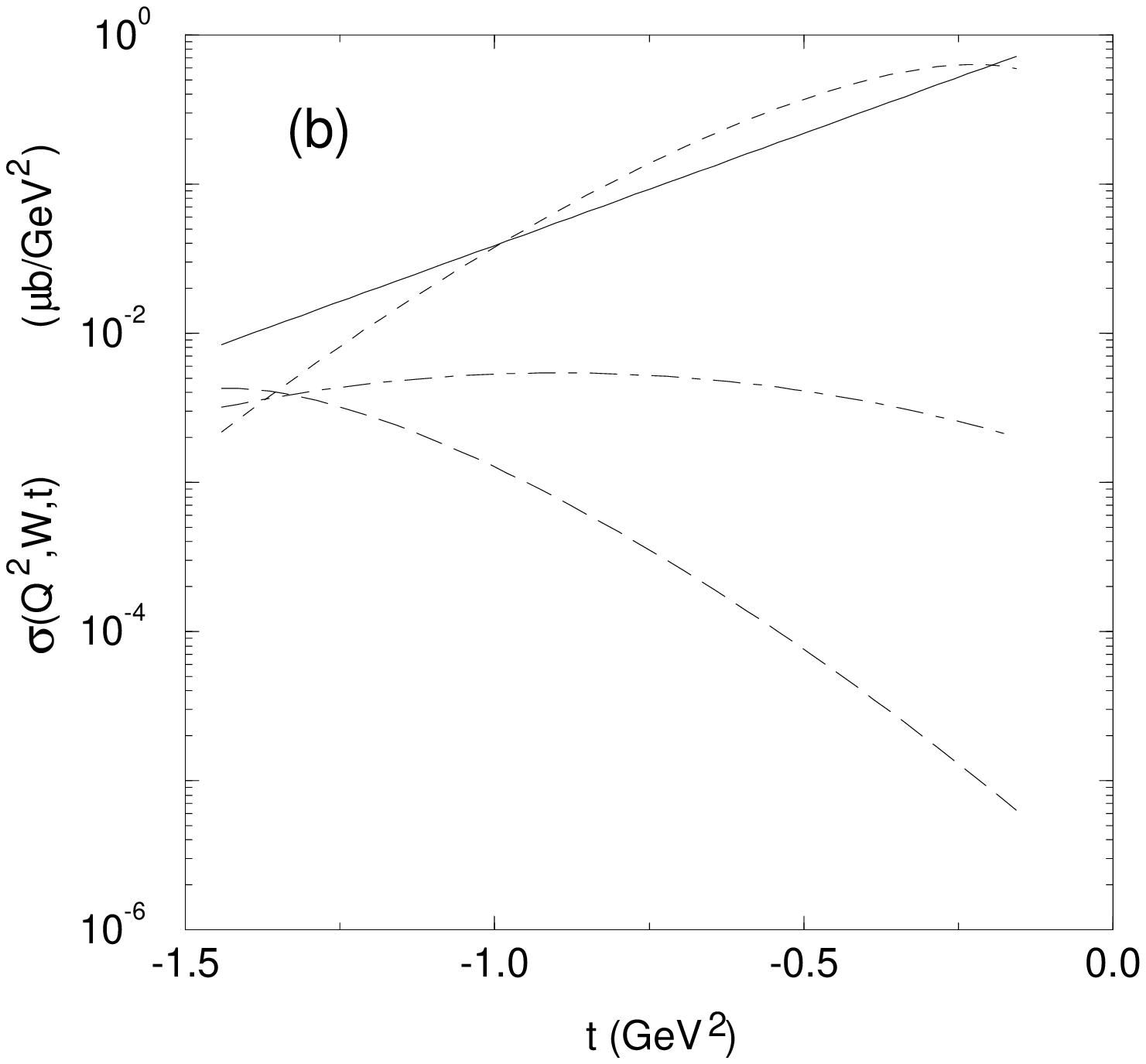}}
\caption{
The $t$-dependence of the cross section $\sigma(Q^2,W,t)$ at $W$=2.1 GeV,
$E_e=11.5$ GeV, and $Q^2=0.02$ GeV$^2$ with (a) $B^2=10$ \% and (b)
$B^2=20$\%. Notations are the same as in Fig.~\protect\ref{nrQ}.}
\label{nrt1}
\end{figure}

\begin{figure}
\vskip 1cm
\centerline{\epsfxsize=0.48\hsize \epsffile{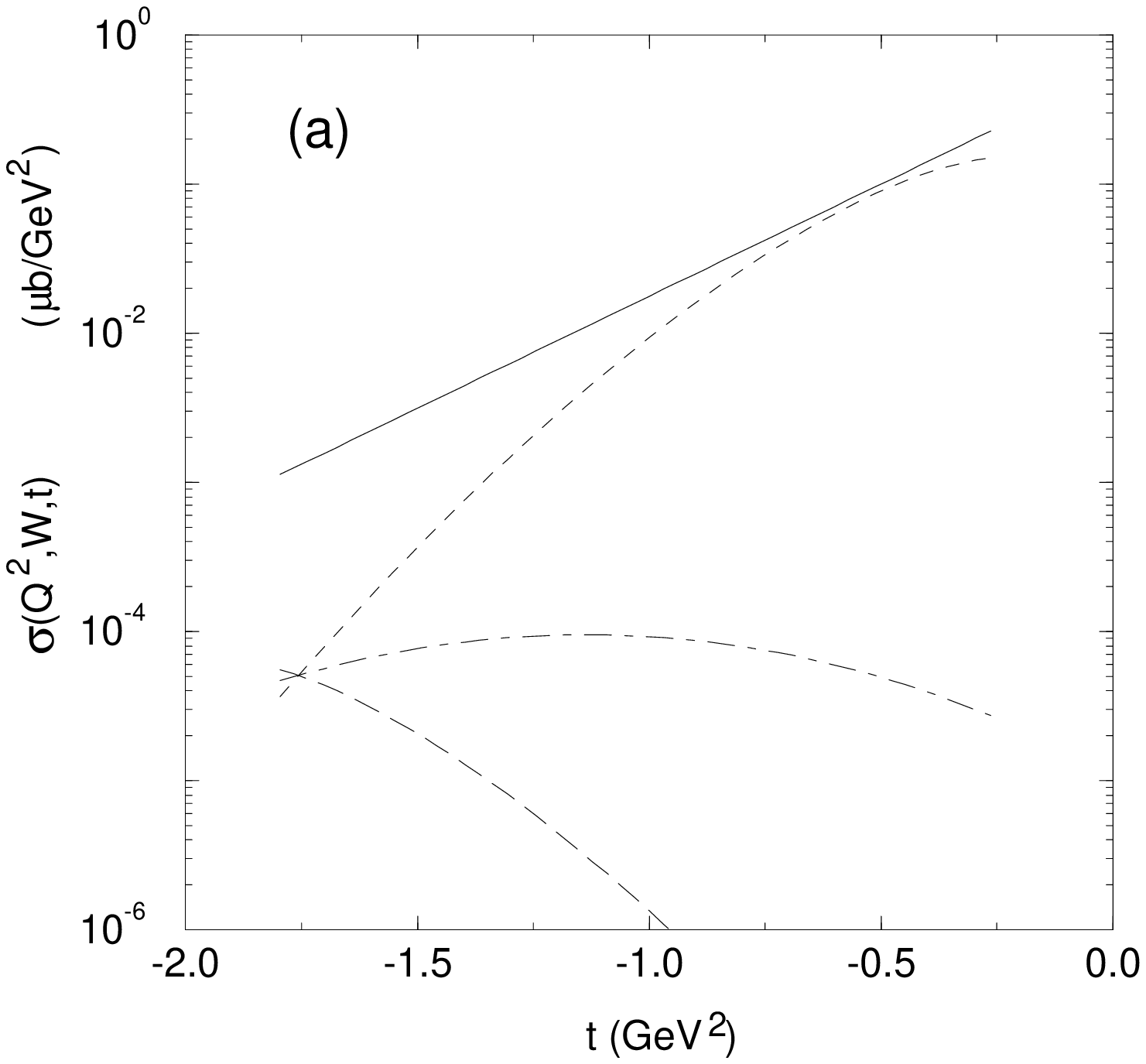}
 \hskip 2cm \epsfxsize=0.48\hsize \epsffile{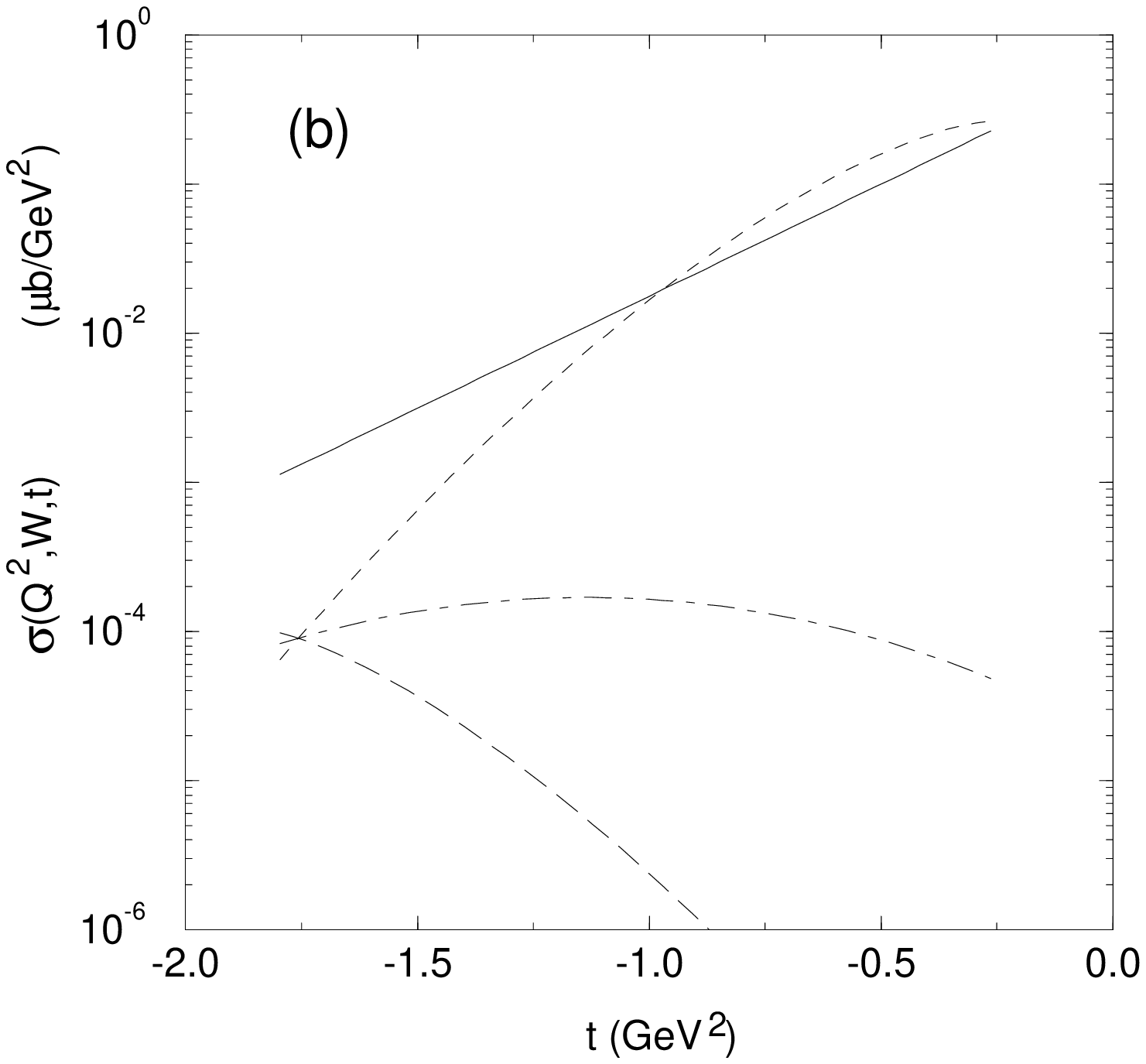}}
\caption{
The $t$-dependence of the cross section $\sigma(Q^2,W,t)$ at $W$=2.1 GeV,
$E_e=11.5$ GeV, and $Q^2=0.5$ GeV$^2$ with (a) $B^2=10$ \% and (b)
$B^2=20$\%. Notations are the same as in Fig.~\protect\ref{nrQ}.}
\label{nrt2}
\end{figure}

The $t$ dependences of the cross section $\sigma(Q^2,W,t)$ at $Q^2 = 0.02$
GeV$^2$ and 0.5 GeV$^2$ are given in Figs. \ref{nrt1} and \ref{nrt2},
respectively. As discussed earlier, the $s \bar s$ knockout cross
section increases with increasing $t$, whereas the $uud$ knockout
cross section decreases.  One can find that at small $Q^2$ and
$t \sim t_{\rm min}$, the three knockout cross sections have the same
order of magnitude. However, except this limited region, the $s \bar s$
knockout process dominates the others.


\section{Relativistic Harmonic Oscillator quark model}

\subsection{The model}

The RHOM first considered in Ref. \cite{FKN} enables one to take
into account the Lorentz-contraction effect of the composite particle
wavefunction. This effect, which is essentially relativistic, becomes
important at large $Q^2$ and provides an explanation of the dipole-like
$Q^2$ dependence of the elastic nucleon form factor. Due to this
advantage, RHOM has been widely used for description of the hadron
properties at large momentum transfers
\cite{FKN,FKR,Lipes,KN,KNO,BOLT}, in spite of some theoretical
difficulties which are inherent in the model \cite{FKR,RPF}.

In RHOM, the spatial motion of a 5-quark system is described by
\begin{eqnarray}
\left( \sum\limits_{i=1}^{5}\Box_i-
\sum\limits_{i\neq j=1}^{5} \kappa_{ij} \left(x_i-x_j\right)^2+V_0 \right)
\Psi=0,
\end{eqnarray}
where $\kappa_{ij}$ and $V_0$ are the usual harmonic oscillator model
parameters, and $x^2 = x_t^2 - \bbox{x}^2$ with $\Box = \partial_t^2 -
\bbox{\nabla}^2$. This equation can be diagonalized using the
relativistic Jacobian coordinates,
\begin{eqnarray}
&&
{\xi_1} = {\textstyle\frac{1}{\sqrt6}}(x_2+x_3-2x_1), \quad
{\xi_2} = {\textstyle\frac{1}{\sqrt2}}(x_3-x_2), \quad
{\rho} = {\textstyle\frac{1}{\sqrt2}}(x_5-x_4), \nonumber\\
&& {\chi} = {\textstyle\sqrt{\frac{2}{15}}}(x_1+x_2+x_3)-
{\textstyle\sqrt{\frac{3}{10}}}(x_4+x_5), \quad
X = {\textstyle\frac{1}{\sqrt5}}\sum\limits_{i=1}^5x_i.
\end{eqnarray}

In the rest frame, $P=(M_0, \bbox{0})$, the ground state spatial
wavefunction is written as \cite{FKN}
\begin{equation}
\Psi^{\rm SP}_{P,\lambda} =
\e^{ - \I M_0 X_t / \sqrt5} \psi^{uud} (\xi_1,\xi_2) \psi^{s\bar s}(\rho)
\psi_\lambda (\chi) =
\e^{ - \I M_0 X_t / \sqrt5} \psi(\xi_1) \psi(\xi_2) \psi(\rho)
\psi_\lambda (\chi),
\end{equation}
and
\begin{eqnarray}
\psi(\upsilon) &=& \left( \frac{\Omega_{\upsilon}}{\pi}\right)
\exp [ - {\textstyle\frac12} \Omega_{\upsilon}(\upsilon_t^2 +
\bbox{\upsilon}^2) ],
\nonumber\\
\psi_{\lambda}(\chi) &=& \sqrt{2\Omega_\chi}\left(
\frac{\Omega_\chi}{\pi}\right) \chi_\lambda \exp [
-{\textstyle\frac12}\omega_\chi(\chi_t^2 +\bbox{\chi}^2) ],
\end{eqnarray}
where $\upsilon = \xi_1$,$\xi_2$,$\rho$ and $\chi_0=\chi_z$ with
$\chi_{\pm 1} = \mp\frac{1}{\sqrt2} (\chi_x \pm \I \chi_y)$. They are
normalized as $\int \d^4 \upsilon |\psi(\upsilon)|^2 = \int \d^4 \chi |
\psi_{\lambda}(\chi)|^2 =1$. The wavefunction with arbitrary momentum
$P$ can be obtained by replacing $-(\upsilon_t^2+\bbox{\upsilon}^2)$ with
\begin{equation}
\upsilon^2 - 2 \left( \frac{P \cdot \upsilon}{M_0} \right)^2.
\end{equation}
Note that in the covariant wavefunction of a state with momentum $P$,
the argument of the Hermite polynomials contains the component of the
space-like four vector $\upsilon_\mu - (P \cdot \upsilon / M^2_0) P_\mu$.
The wavefunctions of the three-quark proton and the $\phi$ meson
in the final state have the similar structure.

In RHOM, the quarks are treated as spinless particles, so we have to
introduce a model for the interaction of the quark spin with the external
electromagnetic field. One of the commonly used methods is to use the
nonrelativistic electromagnetic current \cite{YM,BPCL}. The next order
corrections can be included by Foldy-Wouthuysen transformation, if needed.
Another method is to develop
relativistic generalizations \cite{FKN,FKR,Lipes} by assuming additional
quark spinor structures of the intrinsic wavefunction and by keeping the
additivity of the quark current. The models based on this approach,
therefore, contains some specific assumptions which are not proved yet.
Furthermore, we have to modify the current to satisfy the gauge invariance
condition as in the previous Section. In this work, therefore, by
remembering that the main relativistic effects come from the Lorentz
contraction of the hadron intrinsic wavefunction \cite{FKN}, we use the
following {\em ansatz\/} for the relativistic electromagnetic current,
\begin{mathletters}
\label{RL:current}
\begin{equation}
\hat{J}_\mu^h = \hat{J}^h_{\rm min, \mu} + \hat{J}^h_{\rm mag, \mu},
\end{equation}
with keeping the additivity of the quark current. Here $\hat{J}^h_{\rm min,
\mu}$ is obtained by the minimal substitution and written as
\begin{equation}
\hat{J}^h_{\rm min, \mu} = C \sum_{k=1}^5 e_k \left( p'_{k,\mu} + p_{k,\mu}
\right),
\end{equation}
in momentum space, where $p'_{k,\mu}$ and $p_{k,\mu}$ are the final and
initial 4-momentum of the $k$-th quark, respectively. The normalization
constant $C$ is determined by the condition that $J_0^h = e$ when
$p_k'=p_k$, which gives $C=5/(2M_N)$ in our case. $\hat{J}^h_{\rm mag,
\mu}$ takes part in the interactions of the quark spins with the external
electromagnetic field and has the same form as in NRQM,
\begin{equation}
\hat{J}^h_{\rm mag, 0} = 0, \qquad
\hat{\!\!\bbox{J}}^h_{\rm mag} = \sum_{k=1}^5 \frac{i e_k \mu_k}{2M_N}
\bbox{\sigma}_k \times ( \bbox{p}_k' - \bbox{p}_k ),
\end{equation}
where the magnetic moment $\mu_k$ is defined as $\mu_{u,d} \equiv \mu =
M_N/M_q$ and $\mu_s = M_N/M_s$ as in NRQM.

To determine the model parameters $\Omega_i$, we note that they are
related to the hadron rms radii as
\end{mathletters}
\begin{equation}
\langle r^2_{uud} \rangle = \frac{1}{\Omega_\xi}, \qquad
\langle r^2_{s\bar s} \rangle = \frac34 \frac{1}{\Omega_\rho}.
\end{equation}
We fix $\Omega_\xi$ so as to reproduce the empirical proton magnetic
form factor. Since this model predicts the proton magnetic form factor as
\begin{eqnarray}
G^p_M (Q^2) = \mu (1 + \frac{Q^2}{2 M_N^2} )^{-2}
\exp [- {\textstyle\frac16} \Omega_\xi^{-2}
Q^2 (1 + \frac{Q^2}{2 M_N^2})^{-1} ],
\end{eqnarray}
we can find that $\sqrt{\Omega_\xi}=1.89$ fm$^{-1}$ fits the empirical
dipole formula up to $Q^2 \approx 20$ GeV. This gives the relativistic
scale factor $c^{\rm (RL)}=1.57$. It is interesting to note that this
value is very similar to the nonrelativistic scale factor $c=1.5$
\cite{HKW} used in Sec. IV. We fix $\Omega_\rho$ with $c^{\rm
(RL)}$ as in NRQM calculation. All this process yields
\begin{equation}
\sqrt{\Omega_\xi} = 1.89 \mbox{ fm}^{-1}, \qquad
\sqrt{\Omega_\rho} = 3.02 \mbox{ fm}^{-1}.
\end{equation}
The parameter $\Omega_\chi$ will be discussed in the next subsection.

\subsection{$s \bar s$-knockout}

Because of the structure of the current (\ref{RL:current}), the $s \bar s$
knockout amplitude has the same form as in NRQM,
\begin{equation}
T_{s \bar s}^{\rm (RL)} = - {\textstyle \frac23}
A^* B a_0 \left( \frac{e\mu_s}{2M_N} \right)
\sum_\lambda {\cal I}_\lambda^{s\bar s}
{\textstyle\langle \frac12 \, m_f \, 1 \, \lambda | \frac12 \, m_i \rangle}
(-1)^{m_\phi} ( \bbox{q} \times \bbox{\cal A} )_{-m_\phi}.
\label{Trlss}
\end{equation}
The only difference lies in the spatial overlap integrals, ${\cal
I}_\lambda^{s\bar s}$, defined by
\begin{mathletters} \label{RLSS}
\begin{equation}
{\cal I}_\lambda^{s\bar s} = \tilde\gamma_{s\bar s} \tilde\beta_{s \bar s}
\tilde{\psi}_\lambda^{\rm (RL)},
\end{equation}
where
\begin{eqnarray}
{\tilde\gamma}_{s \bar s} &=& \int \d^4 \xi_1 \d^4 \xi_2 \,
\psi^{uud}_f (p';\xi_1,\xi_2 )
\psi^{uud}_i (p;\xi_1, \xi_2 ),
\\
{\tilde\beta}_{s \bar s} &=& \int \d^4 \rho\,
\e^{\frac{\I}{\sqrt2} q \cdot \rho}
\psi_f^{s \bar s} (q_\phi; \rho) \psi_i^{s \bar s} (p;\rho),
\\
{\tilde \psi}^{\rm (RL)}_{s\bar s, \lambda} (\bbox{p}') &=& \int \d^4 \chi\,
\e^{ \I \chi \cdot (\sqrt{5/6}\, p'-\sqrt{3/10}\, p)}
\psi_\lambda (p;\chi).
\end{eqnarray}
In the relativistic case, the overlap integral $\tilde\beta_{s \bar s}$
depends not only on $\bbox{q}^2(Q^2,W)$ but also on $t$ since the intrinsic
wavefunction depends on $q_\phi$. The integral $\tilde\gamma_{s \bar s}$
is also dependent of $t$ through $p'$ dependence of the intrinsic
wavefunction of the final state.

Then the spin-averaged amplitude squared is
\end{mathletters}
\begin{equation}
\overline{\overline{|T_{s \bar s}^{\rm (RL)}|^2}} =
\frac49
(ABa_0)^2 \frac{e^4}{Q^4}
\tilde\Gamma_{s \bar s} \tilde F_{s \bar s}
\tilde V_{s \bar s} (\bbox{p}')
\cos^2 {\textstyle \frac{\theta}{2}}
\left( g_1^{s \bar s} + g_2^{s \bar s}
\tan^2 {\textstyle \frac{\theta}{2}} \right),
\end{equation}
where
\begin{equation}
g^{s \bar s}_1 = \mu_s^2\eta, \qquad
g^{s \bar s}_2 = 2 \mu_s^2[ \eta + ( \eta + z )^2 ],
\end{equation}
which are the same as $f^{s \bar s}_{1,2}$ of the NRQM. The overlap
integrals can be calculated as described in Appendix. From Eq.
(\ref{RLSS}), we have
\begin{eqnarray}
\tilde\Gamma_{s\bar s} &=& (\tilde{\gamma}_{s\bar s})^2
= ( 1 - \frac{t}{2M_N^2} )^{-4},
\nonumber \\
\tilde F_{s\bar s} &=& (\tilde{\beta}_{s\bar s})^2 =
\left( \frac{M_\phi}{\omega_\phi} \right)^2
\exp[ - q_1^2/(4\Omega_\rho)],
\nonumber \\
\tilde V_{s\bar s} (\bbox{p}')
&=& \frac13 \sum_\lambda | \tilde\psi^{\rm (RL)}_{s\bar s \lambda}
(\bbox{p}') |^2,
\end{eqnarray}
where
\begin{equation}
q_1^2 = \bbox{q}^2 \frac{t}{2M_N \omega_\phi} + \nu^2 \left( 1 -
\frac{\bbox{q}_\phi^2 - \bbox{p}^{\prime 2}}{\omega_\phi \nu} \right).
\label{q12}
\end{equation}
The momentum distribution $\tilde V_{s\bar s} (\bbox{p}')$ is related
to the outgoing proton momentum as in NRQM, but the normalization is
different from $V_{s\bar s} (\bbox{p}')$ of NRQM because of the
normalization condition of the intrinsic wavefunction and an
additional integration over the time component of $\chi$. In NRQM, we
have the usual physical normalization, i.e., one baryon number per unit
volume. We keep this condition in the relativistic case by
renormalizing $\tilde V_{s\bar s} (\bbox{p}')$ as
\begin{equation}
\int \frac{\d \bbox{p}'}{(2\pi)^3} \tilde V_{s\bar s} (\bbox{p}') = 1,
\end{equation}
which gives the final form of $\tilde V_{s\bar s} (\bbox{p}')$ as
\begin{equation}
\frac{1}{(2\pi)^3} \tilde V_{s\bar s} (\bbox{p}') = \frac{ v_{s\bar s}
(\bbox{p}')}{\int \d\bbox{p}' v_{s\bar s} (\bbox{p}')},
\end{equation}
with
\begin{equation}
v_{s\bar s} (\bbox{p}') = \bbox{p}^{\prime 2} \exp\{ - {\textstyle\frac53}
\Omega_\chi^{-1} ( \bbox{p}^{\prime 2} - {\textstyle\frac35} M_N E_p' ) \}.
\end{equation}
We fix the parameter $\Omega_\chi$ so that $v_{s\bar s} (\bbox{p}')$
approaches to the NRQM result with $\bbox{p}^{\prime 2} \ll M_N^2$.
This gives $\Omega_\chi = \frac76 \Omega_\chi^{\rm (NR)}$ by comparing
with the expression given in (\ref{GFV}). With the value determined
in Sec. IV, we get
\begin{equation}
\sqrt{\Omega_\chi} = 2.63 \mbox{ fm}^{-1}.
\end{equation}

\subsection{$uud$-knockout}

By the same way, we can obtain the amplitude of the $uud$ knockout
process as
\begin{equation}
T_{uud}^{\rm (RL)} = \I A^* B a_1 \sum_{j_c,m',m_c,\lambda} b_{2j_c}
{\textstyle
\langle \frac12 \, m' \, 1 \, \lambda | j_c \, m_c \rangle
\langle j_c \, m_c \, 1 \, m_\phi | \frac12 \, m_i \rangle
\langle \frac12 \, m_f | \tilde F_\mu^{uud} | \frac12 \, m' \rangle}
{\cal I}_\lambda^{uud} {\cal A}^\mu.
\label{Trluud}
\end{equation}
With the current (\ref{RL:current}), $\tilde F_\mu^{uud}$ becomes
\begin{equation}
\tilde F^{uud}_0 = ef_0, \qquad
{\tilde{\bbox{F}}}^{uud} = e \bbox{f}_{\rm min} + 
\frac{ \I \mu e}{2M_N} \bbox{\sigma} \times \bbox{q} ,
\label{FFRL}
\end{equation}
where
\begin{eqnarray}
f_0 &=& \frac56 \left( 1 + \frac{E_p' - \omega_\phi}{M_N} +
\frac{2 \bbox{q} \cdot \bbox{p}' }{E_p' M_N} \right), \nonumber \\
\bbox{f}_{\rm min} &=& \frac{5}{6M_N} \left( \bbox{q} - 2 \bbox{q}_\phi +
\frac{2\nu}{E_p'} \bbox{p}' \right).
\end{eqnarray}

As in NRQM, this amplitude does not satisfy the gauge invariance. So we
modify the part of $\bbox{f}_{\rm min}$ which is parallel to $\bbox{q}$ to
satisfy $q^\mu {\tilde F}_\mu^{uud}=0$. This gives
\begin{equation}
\bbox{f}_{\rm min} \to \bbox{f}'_{\rm min} =
\frac{5e}{6M_N} \left\{ - 2 \bbox{q}_\phi + 2
\bbox{q} \cdot \bbox{q}_\phi \frac{\bbox{q}}{\bbox{q}^2} +
\frac{2q_0}{E_p'} \left( \bbox{p}' - \bbox{q} \cdot \bbox{p}'
\frac{\bbox{q}}{\bbox{q}^2} \right) \right\} + f_0 q_0
\frac{\bbox{q}}{\bbox{q}^2}.
\end{equation}

The overlap integral ${\cal I}_\lambda^{uud}$ is defined as
\begin{mathletters}
\begin{equation}
{\cal I}_\lambda^{uud} = \tilde\gamma_{uud} \tilde\beta_{uud}
\tilde \psi^{\rm (RL)}_{uud,\lambda} (\bbox{q}_\phi),
\end{equation}
where
\begin{eqnarray}
\tilde\gamma_{uud} &=& \int \d^4 \rho \, \psi_f^{s \bar s} (q_\phi;\rho)
\psi_i^{s \bar s} (p;\rho),
\\
\tilde\beta_{uud} &=& \int \d^4 \xi_1 \d^4 \xi_2 \, \e^{\I
\sqrt{\frac23} q \cdot \xi_1 } \psi_f^{uud}
(p'; \xi_1, \xi_2) \psi_i^{uud} (p; \xi_1, \xi_2),
\\
\tilde\psi^{\rm (RL)}_{uud,\lambda} (\bbox{q}_\phi) &=& \int \d^4 \chi
\, \e^{\I \chi \cdot ( \sqrt{5/12}\, p - \sqrt{5/6}\, q_\phi)}
\psi_\lambda (p; \chi).
\end{eqnarray}

The squared matrix element is, therefore,
\end{mathletters}
\begin{equation}
\overline{\overline{|T^{\rm (RL)}_{uud}|^2 }}
= ( ABa_1)^2 \frac{e^4}{Q^4}
\tilde\Gamma_{uud} \tilde F_{uud}(\bbox{q}) \tilde V_{uud} (\bbox{q}_\phi)
\cos^2 {\textstyle \frac{\theta}{2}}
\left( g_1^{uud} + g_2^{uud} \sin^2 {\textstyle \frac{\theta}{2}}
\right),
\label{T2}
\end{equation}
where
\begin{eqnarray}
g_1^{uud} &=& \eta \mu^2 \left\{ 1 + \frac{f_0^2}{\mu^2}
\frac{\eta}{ [\eta + (\eta+z)^2]^2 } + \frac{25}{9} \frac{c}{\mu^2}
\left( 1+ \frac{2(\eta+z)}{1+2y} \right) \right\},
\nonumber \\
g_2^{uud} &=& 2 \mu^2 [ \eta + (\eta+z)^2 ] \left\{ 1 + 
\frac{25}{9} \frac{c}{\mu^2} \left( 1+ \frac{2(\eta+z)}{1+2y} \right)
\right\},
\label{g12uud}
\end{eqnarray}
with $c$ defined in (\ref{def:c}). By analyzing the numerical values of
$g_{1,2}^{uud}$ we can find that $g_{1,2}^{uud} \approx f_{1,2}^{uud}$.
This means that {\it the magnetic part of the e.m. current gives
the main contribution as in NRQM.}

By making use of the formulas given in Appendix, the overlap integrals
can be calculated as
\begin{eqnarray}
\tilde\Gamma_{uud} &=& (\tilde{\gamma}_{uud})^2
= \left( \frac{M_\phi}{\omega_\phi} \right)^2,
\nonumber \\
\tilde F_{uud} &=& (\tilde{\beta}_{uud})^2 =
( 1 - \frac{t}{2M_N^2} )^{-4}
\exp[ - q_2^2/(3\Omega_\xi)],
\end{eqnarray}
where
\begin{equation}
q_2^2 = \bbox{q}^2 \left( 1 - \frac{\nu}{E_p'} \right) +
\nu^2 \left( 1 - \frac{\bbox{p}^{\prime 2} - \bbox{q}_\phi^2}{\nu E_p'}
\right).
\label{q22}
\end{equation}
As in the $s\bar s$ knockout case, they depend on $t$ as well as on $Q^2$
and $W$. In Eq. (\ref{T2}), $\tilde V_{uud} (\bbox{q}_\phi)$ is defined as
$\frac13 \sum_\lambda | \tilde\psi^{\rm (RL)}_{uud,\lambda} |^2$ and is
normalized
as in $\tilde V_{s\bar s} (\bbox{p}')$, i.e.,
\begin{equation}
\frac{1}{(2\pi)^3} \tilde V_{uud} (\bbox{q}_\phi) = \frac{ v_{uud}
(\bbox{ q}_\phi)}{\int \d\bbox{q}_\phi v_{uud} (\bbox{q}_\phi)},
\end{equation}
with
\begin{equation}
v_{uud} (\bbox{q}_\phi) = \bbox{q}^2_\phi \exp\{ - {\textstyle\frac53}
\Omega_\chi^{-1} ( \bbox{q}^2_\phi - {\textstyle\frac25} M_N \omega_\phi )
\}.
\end{equation}
Note that it contains factor $\frac25$ while $v_{s\bar s}$ has $\frac35$.

\subsection{interference}

The interference between the two knockout processes are obtained as
\begin{equation}
\overline{\overline{|T_{int}^{\rm (RL)}|^2}} =
c_{\rm int} (AB)^2 a_0 a_1 \frac{e^4}{Q^4}
\tilde\Gamma_{\rm int} \tilde F_{\rm int}
\tilde V_{\rm int} (\bbox{p}' , \bbox{q}_\phi)
\cos^2 {\textstyle \frac{\theta}{2}}
\left( g_1^{\rm int} + g_2^{\rm int}
\tan^2 \textstyle{\frac{\theta}{2}} \right),
\end{equation}
where
\begin{eqnarray}
c_{\rm int} &=& \frac{4}{9\sqrt3} (b_1 + 2 \sqrt2 b_3 ),
\nonumber \\
g_1^{\rm int} &=& \mu\mu_s \eta = f_1^{\rm int},
\nonumber \\
g_2^{\rm int} &=& 2\mu\mu_s [ \eta + (\eta+z)^2] = f_2^{\rm int},
\end{eqnarray}
and
\begin{eqnarray}
\tilde\Gamma_{\rm int} &=& \tilde\gamma_{s \bar s} \tilde\gamma_{uud} 
= \frac{M_\phi}{\omega_\phi} (1- \frac{t}{2M_N^2} )^{-2},
\nonumber \\
\tilde F_{\rm int} &=& \beta_{s \bar s} \beta_{uud}
= \tilde\Gamma_{\rm int} \exp\left( - \frac{1}{8\Omega_\rho} q_1^2 -
\frac{1}{6\Omega_\xi} q_2^2 \right),
\nonumber \\
\tilde V_{\rm int} ( \bbox{p}' , \bbox{q}_\phi) &=& -N \bbox{p}' \cdot
\bbox{q}_\phi \exp\left[ - \frac{5}{6\Omega_\chi}
\left( \bbox{p}^{\prime 2} + \bbox{q}_\phi^2 - \frac35 E_p' M_N -
\frac25 \omega_\phi M_N \right) \right\},
\end{eqnarray}
where $N$ is the normalization constant which can be derived from the
normalization conditions of $\tilde V_{s\bar s}$ and $\tilde V_{uud}$.
As in NRQM, $\tilde V_{\rm int}$ has overall $(-)$ sign.

\subsection{results}

The main difference between the NRQM and the RHOM lies in the overlap
integrals due to the similar structure of the electromagnetic currents.
In RHOM, the overlap integrals, $\tilde\Gamma_{s\bar s}$,
$\tilde\Gamma_{uud}$ are smaller than 1 ($=\Gamma_{s\bar s, \, uud}$)
over the whole area of $t$.  This effect is stronger in
$\tilde\Gamma_{s\bar s}$ than in $\tilde\Gamma_{uud}$ and becomes
larger at small (large) $t$ for $\tilde\Gamma_{s\bar s}$
($\tilde\Gamma_{uud}$). This holds also for $\tilde V_{s\bar s}$ and
$\tilde V_{uud}$. In this case, however, the suppression is stronger
in $\tilde V_{uud}$ than in $\tilde V_{s\bar s}$. These bring some
suppressions of the RHOM amplitudes.

\begin{figure}
\vskip 1cm
\centerline{\epsfxsize=0.48\hsize \epsffile{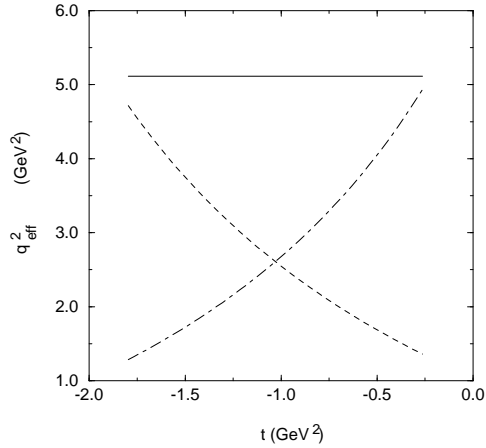}}
\caption{ The $t$ dependence of $q_1^2$ and $q_2^2$ in the RHOM
overlap integrals at $Q^2=0.5$ GeV$^2$. The solid line is the value
of $\bbox{q}^2$, and the dotted and dot-dashed lines are $q_1^2$ and
$q_2^2$, respectively.}
\label{fig:qef}
\end{figure}

However, such suppressions are of order of 10 at most and they are
compensated and dominated by strong enhancement of
$\tilde F_{s\bar s,\,uud}$ that contain the wavefunctions of the
struck quarks. The $Q^2$ dependence of the amplitudes is mainly
determined by $\tilde F_{s\bar s,\,uud}$. In RHOM, they depend on the
effective momentum transfers, $q_1^2$ and $q_2^2$, as defined in
(\ref{q12}) and (\ref{q22}). Figure \ref{fig:qef} shows the
$t$-dependence of $q_{1,2}^2$ at $Q^2=0.5$ GeV$^2$. One can find that
$q_{1,2}^2$ are always smaller than $\bbox{q}^2$ and the suppression
becomes larger at large (small) $t$ for $q_1^2$ ($q_2^2$). This results in
enhancement of $\tilde F_{s\bar s}$ and $\tilde F_{uud}$, which is strong
enough to dominate over the suppression of the other overlap integrals. As
a result, the RHOM amplitudes are enhanced as a whole.

\begin{figure}
\vskip 1cm
\centerline{\epsfxsize=0.48\hsize \epsffile{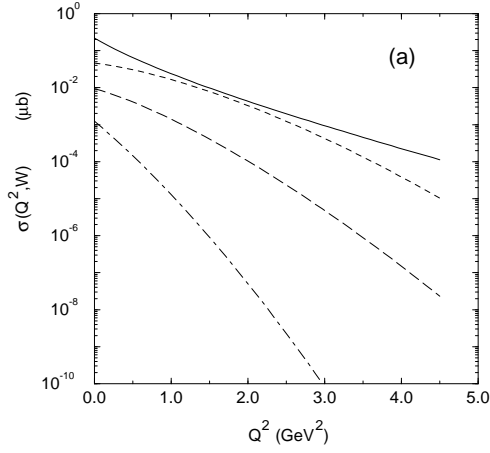}
 \hskip 2cm \epsfxsize=0.48\hsize \epsffile{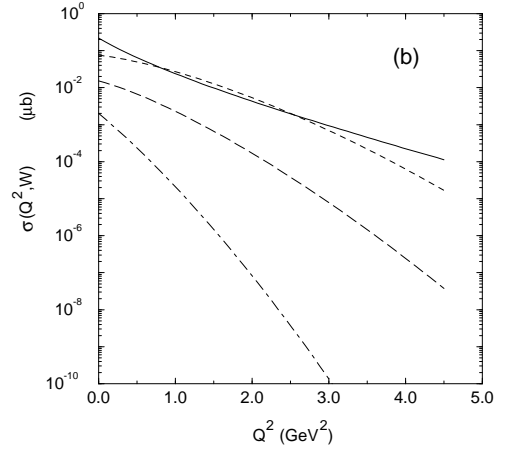}}
\caption{
The $Q^2$-dependence of the cross section
$\sigma(Q^2,W)$ at $W$=2.1 GeV and $E_e=11.5$ GeV in RHOM with (a)
$B^2=3$ \% and (b) $B^2=5$\%. Notations are the same as in
Fig.~\protect\ref{nrQ}.}
\label{rhQ}
\end{figure}

Displayed in Fig. \ref{rhQ} is the RHOM predictions of $\sigma(Q^2,W)$ with
$B^2=3$ and 5\%. We find that the RHOM prediction exceeds the NRQM ones and
the difference reaches several orders of magnitude. This is mainly because
of the new functional dependence of $\tilde F_{s\bar s,\,uud}$ in RHOM.
This relativistic modification of form factors is more crucial for $\tilde
F_{uud}$ due to the fact that the rms radius of $uud$ cluster is larger than
that of $s\bar s$ cluster. Only at small $Q^2$ the RHOM result is close to
NRQM one. As in NRQM, we can find that the $s\bar s$ knockout process is
the dominant one. The $uud$ knockout and the interference are small.
As a result, we can find that {\it because of the strong enhancement of the
form factors the cross section of the knockout process is comparable to that
of VDM only with $B^2=3 \sim 5$\%.}

\begin{figure}
\vskip 1cm
\centerline{\epsfxsize=0.48\hsize \epsffile{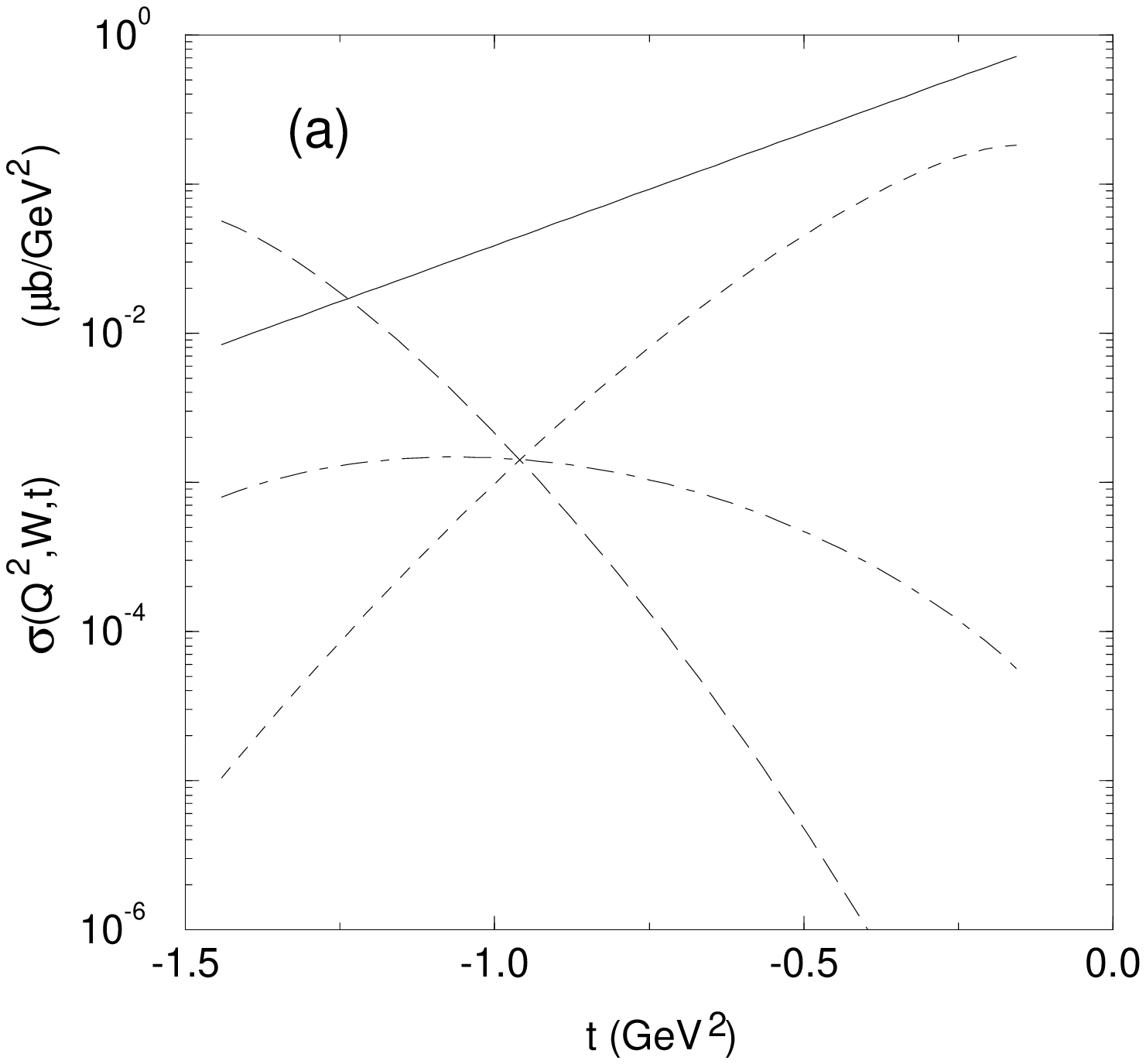}
 \hskip 2cm \epsfxsize=0.48\hsize \epsffile{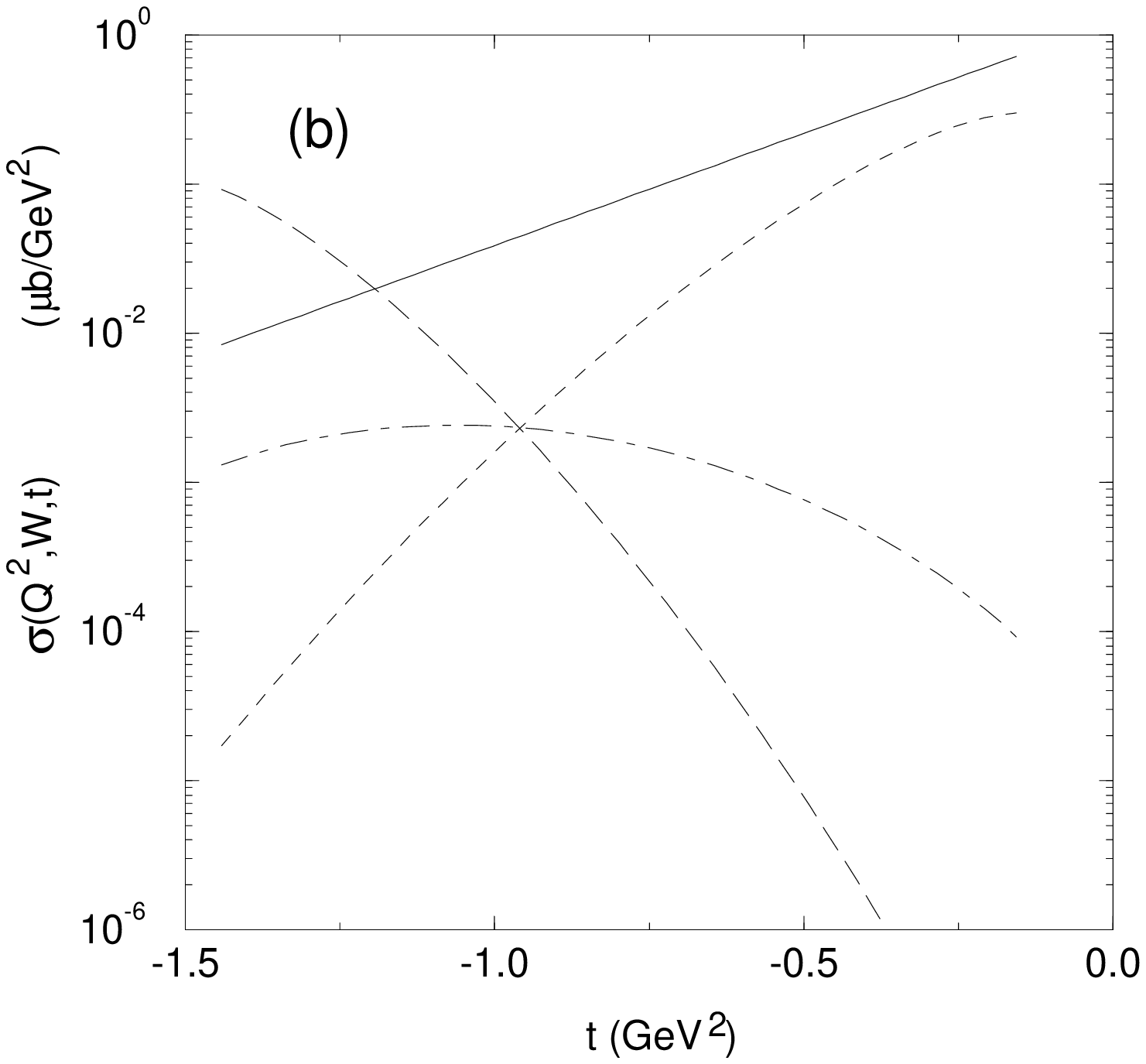}}
\caption{
The $t$-dependence of the cross section $\sigma(Q^2,W,t)$ at $W$=2.1 GeV,
$E_e=11.5$ GeV, and $Q^2=0.02$ GeV$^2$ with (a) $B^2=3$ \% and (b)
$B^2=5$\%. Notations are the same as in Fig.~\protect\ref{nrQ}.}
\label{rht1}
\end{figure}

\begin{figure}
\vskip 1cm
\centerline{\epsfxsize=0.48\hsize \epsffile{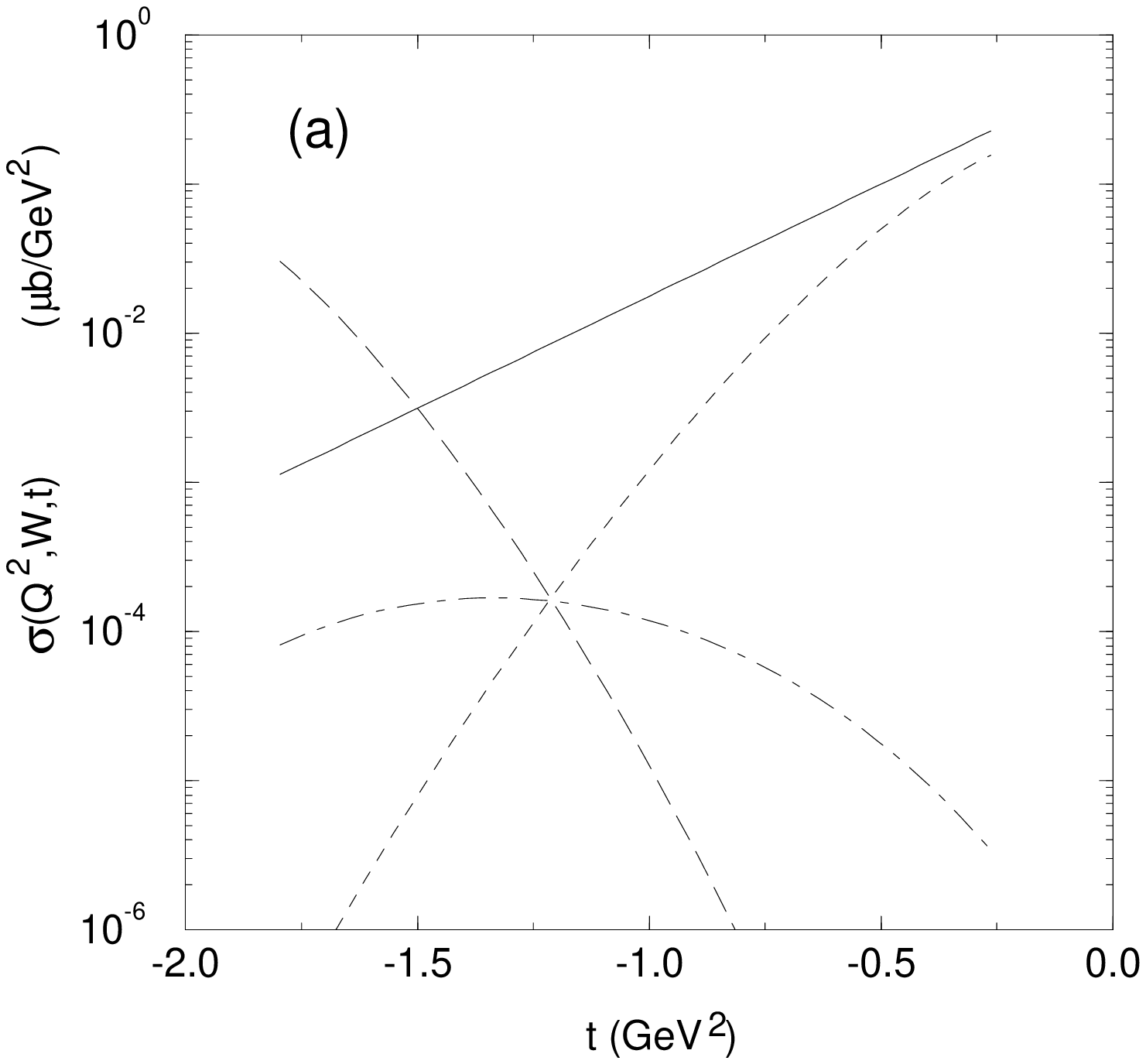}
 \hskip 2cm \epsfxsize=0.48\hsize \epsffile{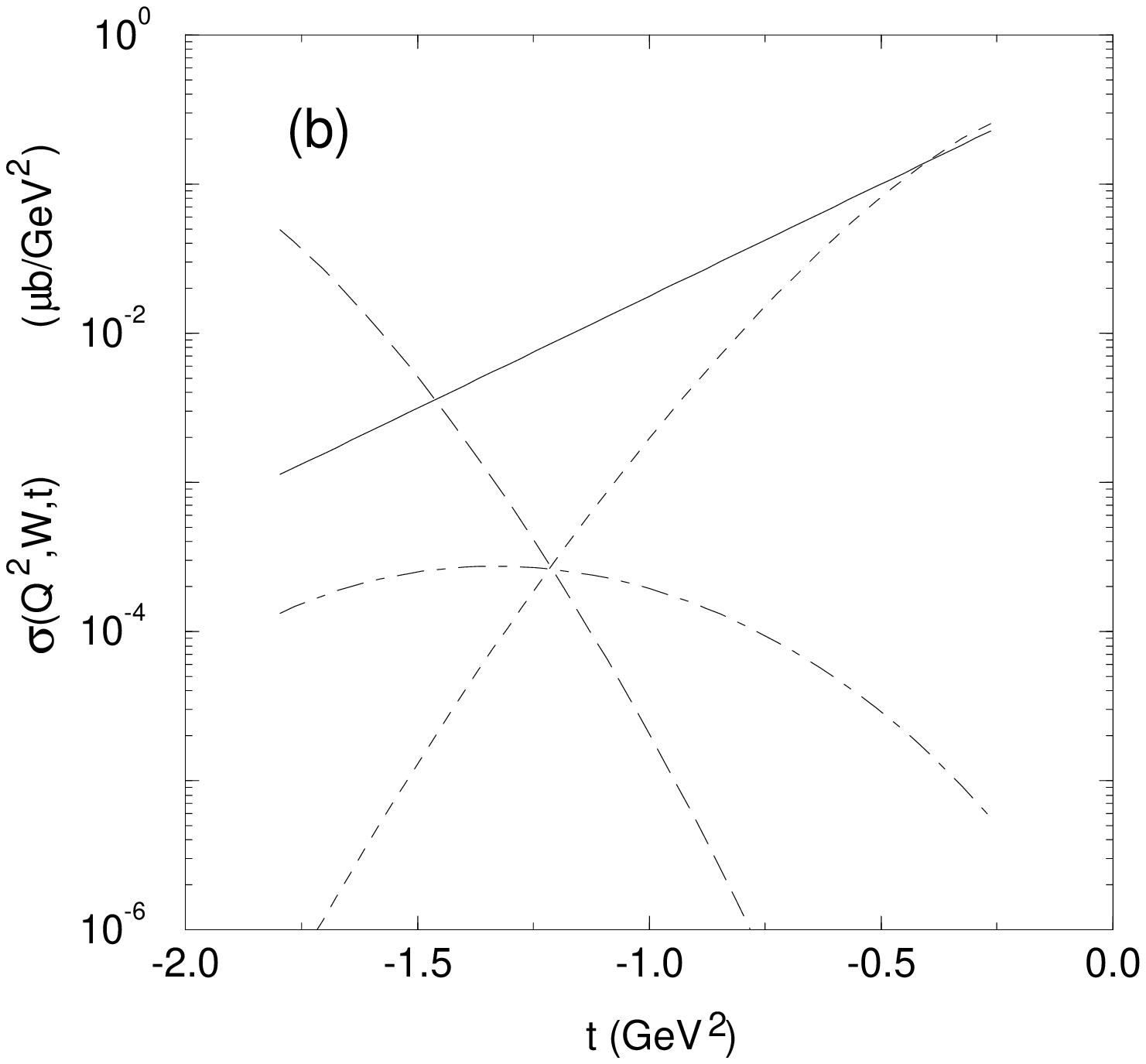}}
\caption{
The $t$-dependence of the cross section $\sigma(Q^2,W,t)$ at $W$=2.1 GeV,
$E_e=11.5$ GeV, and $Q^2=0.5$ GeV$^2$ with (a) $B^2=3$ \% and (b)
$B^2=5$\%. Notations are the same as in Fig.~\protect\ref{nrQ}.}
\label{rht2}
\end{figure}

\begin{figure}
\vskip 1cm
\centerline{\epsfxsize=0.48\hsize \epsffile{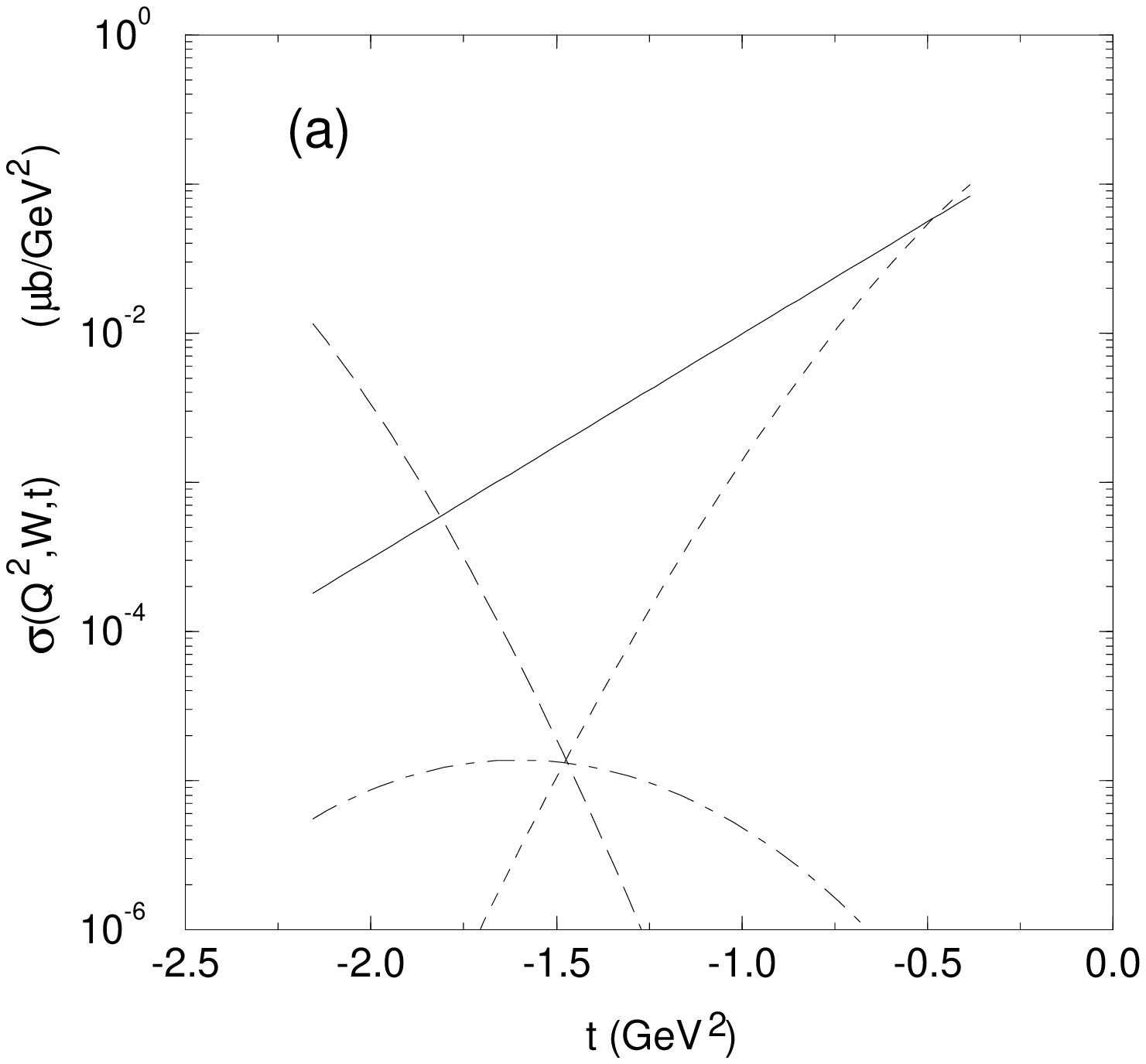}
 \hskip 2cm \epsfxsize=0.48\hsize \epsffile{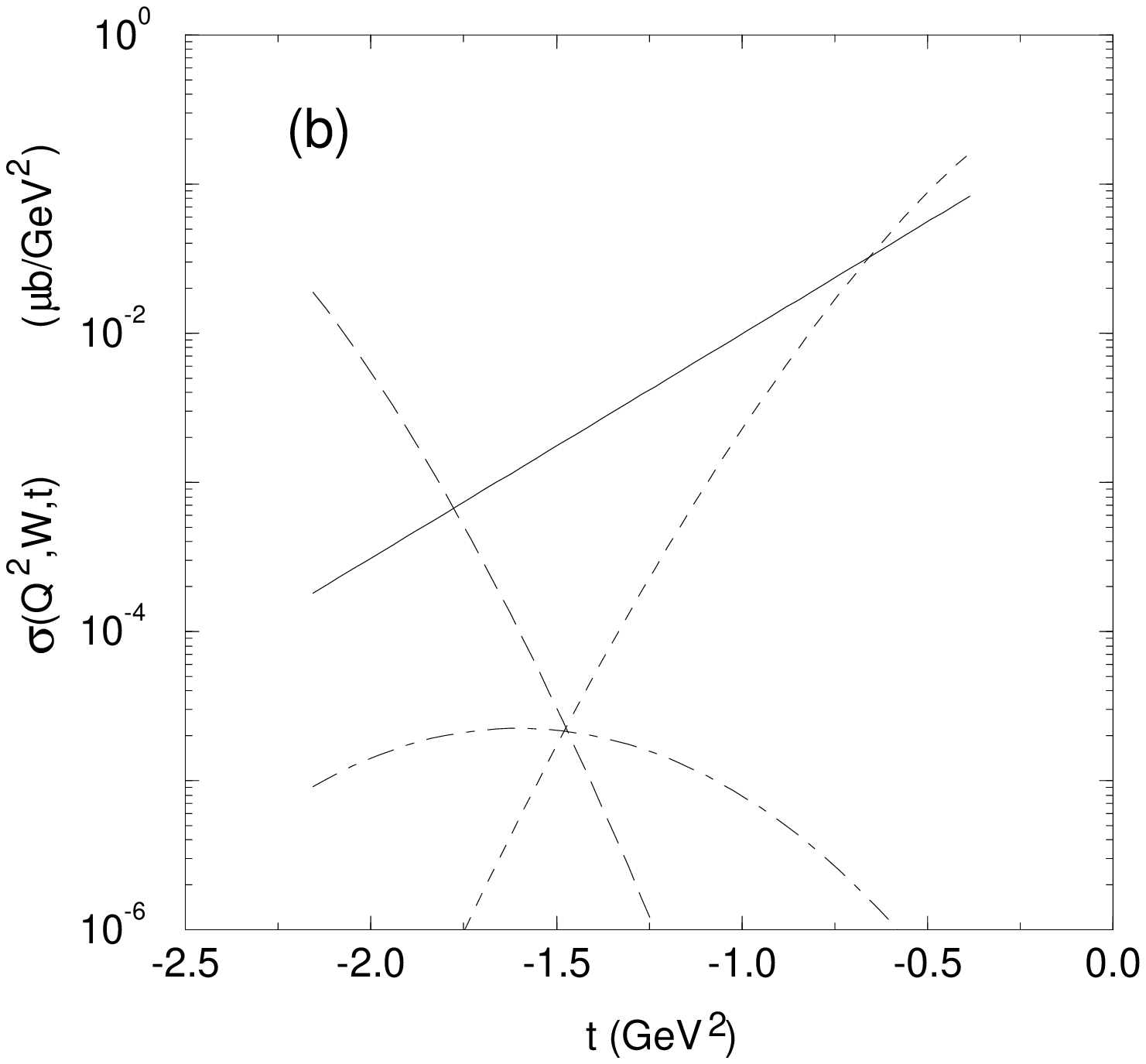}}
\caption{
The $t$-dependence of the cross section $\sigma(Q^2,W,t)$ at $W$=2.1 GeV,
$E_e=11.5$ GeV, and $Q^2=1.0$ GeV$^2$ with (a) $B^2=3$ \% and (b)
$B^2=5$\%. Notations are the same as in Fig.~\protect\ref{nrQ}.}
\label{rht3}
\end{figure}

In Figs. \ref{rht1}--\ref{rht3} we present the $t$-dependence of the
cross section $\sigma(Q^2,W,t)$ at $Q^2=0.02$ GeV$^2$, 0.5 GeV$^2$,
and 1.0 GeV$^2$, respectively. The RHOM gives strong enhancement of the
$uud$ knockout cross sections.  Because of the suppressions of form
factors at small $t$, the $s \bar s$ knockout cross section is smaller
than the NRQM one at small $t$ and $Q^2$, where the enhancement of
$\tilde F_{s\bar s}$ is small. However as $Q^2$ increases, we can see
that the $s \bar s$ knockout cross section is comparable to VDM one at
large $t$. {\it At small $t$, the $uud$ knockout is the dominant one\/}
and exceeds even the VDM prediction near $t_{\rm min}$. The effect of
the interference is small and is suppressed as $Q^2$ increases.

Note that our result is based mainly on the Lorentz contraction effect of
the wavefunction and is not sensitive to the models of hadron
electromagnetic current in the relativistic model.
For instance, if we compare $\sigma(Q^2,W)$ of the $s \bar s$ knockout
with that of Ref. \cite{TY} which is obtained with relativistic hadron
electromagnetic current of Ref. \cite{FKN}, the main difference between
the two comes from an additional kinematical factor of the amplitude.
However, the two predictions are close to each other within the accuracy
of the model.


\section{Strangeness coefficients of the proton in quark model}

We have seen that the $\phi$ electroproduction cross sections are very
sensitive to the strangeness coefficients $B$, $a_0$, and $a_1$ of the
proton wavefunction (\ref{protonwf}). The coefficient $Ba_0$ is directly
related to the $s\bar s$-knockout process and $Ba_1$ to the
$uud$-knockout. The purpose of this study is to determine theoretical upper
bounds of these coefficients by comparing the cross sections of the
knockout process for the $\phi$ electroproduction with the experimental
data.  Because of the shortage of the data, however, it will be
interesting to estimate them in other phenomenologically successful
models at the present stage. In this Section, we briefly discuss a way
to extract the strangeness coefficients of the proton wavefunction,
i.e., $B^2$, from an extended quark model using some results of the
cloudy bag model.

When we take into account ($3q$)($q\bar q$) admixture in the proton
wavefunction as well as the conventional $3q$-structure, we can write
the physical proton wavefunction as
\begin{equation}
\mid \Psi_p \rangle = A_0 \mid \psi_{p_0} \rangle + \sum_{B,M} A_{_{BM}}
{\cal A} \mid \psi_{_B} \psi_{_M} \varphi_{_{BM}} \rangle,
\label{p-wf}
\end{equation}
where $\psi_{p_0}$ is the conventional 3-quark proton wavefunction,
$\psi_{_B}$ and $\psi_{_M}$ are the baryonic core and mesonic cloud
wavefunctions, respectively, with $\varphi_{_{BM}}$ describing their
relative motion. ${\cal A}$ is the antisymmetrizer which guarantees
the symmetry property of $\Psi_p$ under interchange of quark labels.
Then the quantities of our interests can be obtained as
\begin{equation}
Ba_0 = \langle p_0 \eta_s \varphi_{p_0 \eta_s} \mid \Psi_p
\rangle,
\quad
Ba_1 = \langle p_0 \phi_s \varphi_{p_0 \phi_s} \mid \Psi_p
\rangle.
\end{equation}
Hereafter, we use $\eta_s$ and $\phi_s$ for $[s\bar s]^0$ and $[s\bar
s]^1$ of (\ref{protonwf}), respectively. Although the core and cloud
wavefunctions, $\psi_{_B}$ and $\psi_{_M}$ in (\ref{p-wf}), can be any of
the usual color-singlet baryon and meson or their resonances \cite{JL},
we expect that the main contribution comes from the combinations of the
ground states \cite{KH} on energetic grounds. In addition, there can be
hidden color combinations between them; i.e., $\psi_{_B}$ can be a
wavefunction of a non-color-singlet particle which form a totally
color-singlet proton state with a non-color-singlet meson cloud.
However, in the nonrelativistic quark model \cite{FH}, it was shown
that the hidden color combinations contribute little to the physical
proton wavefunction. Therefore, we neglect these combinations in our
simple calculation. $A_0$, $A_{BM}$, and $\varphi_{BM}$ may
be found from phenomenological models, e.g., by diagonalizing the
nonrelativistic quark model Hamiltonian \cite{FH} or from the chiral bag
model. In our simple calculation, we will use the cloudy
bag model \cite{Cloudy}, which has an advantage that the
probabilities of the meson clouds can be read easily.

\begin{figure}
\vskip 1cm
\centerline{\epsfxsize=0.48\hsize \epsffile{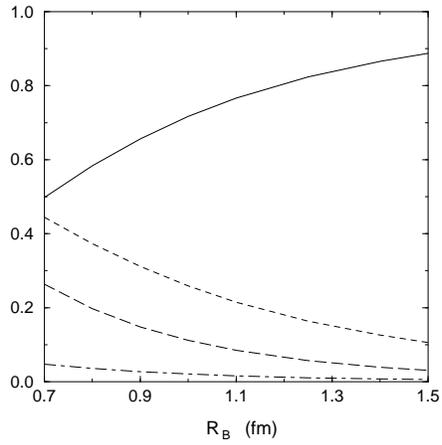}}
\caption{
Prediction of the cloudy bag model on the probabilities of meson clouds
in proton as a function of bag radius $R_B$ (in fm).
The solid line represents the probability of the bare proton, the dotted
line the $\pi$ cloud, the dashed (dot-dashed) line the $K$ ($\eta$) cloud.
The dashed and dot-dashed lines are exaggerated by 5 and 10 times,
respectively.}
\label{fig:cbm}
\end{figure}

As in most versions of the cloudy bag model, we will consider only the
pseudoscalar octet meson clouds. To this end, we extend the work of
Ref. \cite{CBM} by including the $\eta$ meson cloud as well. The
probabilities of meson clouds are given in Fig. \ref{fig:cbm} as a
function of the bag radius $R_B$. We can verify that the meson cloud
effects are important at small bag radius and the roles of the $K$
and $\eta$ clouds are much smaller than that of the $\pi$ cloud; The
probability of $K$ cloud is less than 4\% and that of $\eta$ cloud is
less than 0.5\% in the region of $R_B \geq 0.7$ fm.

Our method to estimate the strangeness coefficients is very similar to
the determination of the 6-$q$ configuration of the deuteron wavefunction
\cite{BOLT,MS}. As a specific example, let us consider the $\Lambda K^+$
combination of the proton wavefunction, which can be written as
\begin{equation}
\psi_{5q} (\Lambda K)
    = \frac{1}{N} ( 1 - \sum_{{i=1,2,3}\atop{j=4,5}} {\hat P}_{ij} )
      \psi_\Lambda \psi_{K^+} \varphi_{\Lambda K^+},
\label{psi5q}
\end{equation}
where $N$ is the normalization constant and ${\hat P}_{ij}$ is the
permutation operator exchanging the color, spin-flavor, and spatial
coordinates of quarks in the core and the cloud;
${\hat P}_{ij} = {\hat P}^C_{ij} {\hat P}^{SF}_{ij} {\hat P}^X_{ij}$.
So the probability of $(B' M')$ configuration in
$\psi_{5q} (\Lambda K)$ of (\ref{psi5q}) is
\begin{eqnarray}
{\cal P} ( B' M' ) &=&
 \frac{ \mid \langle B' M' \varphi_{B'M'} \mid \psi_{5q} (\Lambda K)
        \rangle \mid^2 }
      { \sum_{B,M} \mid \langle B M \varphi_{BM} \mid
        \psi_{5q} (\Lambda K) \rangle \mid^2 }
\nonumber \\
&=&
 \frac{ \mid \delta_{B'\Lambda} \delta_{M'K^+}
        - 6 \langle B' M' \varphi_{B'M'} \mid {\hat P}_{34} \mid
        \Lambda K^+ \varphi_{\Lambda K^+} \rangle \mid^2 }
      { \sum_{B,M} \{ \mid \delta_{B\Lambda} \delta_{MK^+} - 6
        \langle B M \varphi_{BM} \mid {\hat P}_{34} \mid \Lambda K^+
        \varphi_{\Lambda K^+} \rangle \mid^2 \} }.
\end{eqnarray}
Here we shall consider the case that $(B,M)$=$(\Lambda
K^+)$,$(p_0,\eta_s)$,$(p_0,\phi)$; the contributions of the other
clusters are found to be negligible.

Then it is straightforward to calculate the color exchange term as
\begin{equation}
\langle {\hat P}^C_{34} \rangle = \textstyle \frac16,
\end{equation}
for all configurations. From the spin-flavor wavefunction of the clusters
\cite{Close}, we obtained
\begin{mathletters}
\begin{eqnarray}
\langle \Lambda K^+ \mid {\hat P}^{SF}_{34} \mid \Lambda K^+ \rangle &=&
\textstyle \frac{1}{12}, \\
\langle p_0 \eta_s \mid {\hat P}^{SF}_{34} \mid \Lambda K^+ \rangle &=&
\textstyle \frac{1}{4\sqrt6},
\end{eqnarray}
after some exercises. For the $p_0 \phi$ configuration, we have two types
of them because $\phi$ is a vector meson. When we write $\bbox{J}_c =
\bbox{J}_B + \bbox{L}$ which forms the total spin \bbox{J} ($j=\frac12$)
with the meson spin $\bbox{J}_M$, then $j_c$ can be either $\frac12$ or
$\frac32$. We distinguish the two by
representing $\phi_s^1$ and $\phi_s^3$ for $[s\bar s]^1$ cluster with
$j_c=\frac12$ and $j_c=\frac32$, respectively, which leads us to
\begin{eqnarray}
\langle p_0 \phi_s^1 \mid {\hat P}^{SF}_{34} \mid \Lambda K^+ \rangle &=&
\textstyle \frac{1}{4\sqrt2}, \\
\langle p_0 \phi_s^3 \mid {\hat P}^{SF}_{34} \mid \Lambda K^+ \rangle &=&
0.
\end{eqnarray}
Note that the recoupling coefficient between the $p_0\phi_s^3$ and $\Lambda
K^+$ is zero because of the structure of the spin-flavor wavefunctions.

For internal wavefunctions of the clusters, we use those of the
harmonic oscillator potential model:
\end{mathletters}
\begin{equation}
\psi^B_{\text{int}} = \left( \frac{\Omega_B}{\pi} \right)^{3/2}
     \exp [ -{\textstyle\frac12} \Omega_B
             ( \bbox{\xi}_1^2 + \bbox{\xi}_2^2 ) ],
\qquad
\psi^{M}_{\text{int}} = \left( \frac{\Omega_M}{\pi} \right)^{3/4}
     \exp ( -{\textstyle\frac12} \Omega_M
              \bbox{\rho}^2 ),
\end{equation}
in terms of the Jacobian coordinates defined in (\ref{Jac}). For simplicity,
we assume that $M_u$=$M_d$=$M_s$ and the same $\Omega_B$ ($\Omega_M$) for
baryon (meson) clusters. After performing ${\hat P}^X_{34}$, the Jacobian
coordinates transform as
\begin{eqnarray}
\bbox{\xi}_1 &\to& \bbox{\xi}_1' = {\textstyle \frac13} \bbox{\xi}_1
    + {\textstyle \frac{2}{\sqrt6}} \bbox{\chi}
    - {\textstyle \frac{1}{\sqrt6}} \bbox{\rho}, \nonumber \\
\bbox{\xi}_2 &\to& \bbox{\xi}_2 = \bbox{\xi}_2, \nonumber \\
\bbox{\rho} &\to& \bbox{\rho}' = \bbox{\chi}
    - {\textstyle \sqrt{\frac23}} \bbox{\xi}_1
    + {\textstyle \frac12} \bbox{\rho}, \nonumber \\
\bbox{\chi} &\to& \bbox{\chi}' = {\textstyle \frac16} \bbox{\chi}
    + {\textstyle \frac{5}{12}} \bbox{\rho}
    + {\textstyle \frac56 \sqrt{\frac23}} \bbox{\xi}_1.
\end{eqnarray}
Then we have
\begin{equation}
\langle B' M' \varphi_{B'M'} \mid {\hat P}^X_{34} \mid
              \Lambda K^+ \varphi_{\Lambda K^+} \rangle
= \int \d \bbox{\chi} \d \bbox{\chi}' \varphi_{B' M'} (\bbox{\chi})
  K ( \bbox{\chi}, \bbox{\chi}' )
  \varphi_{\Lambda K^+} (\bbox{\chi}'),
\label{P34X}
\end{equation}
where
\begin{equation}
K (\bbox{\chi}, \bbox{\chi}') = \left( \frac{12}{5} \right)^3
  \left( \frac{3 \Omega_M}{(3+8g)\pi} \right)^{3/2}
  \exp [ - h_1 ( \bbox{\chi}^2 + \bbox{\chi}^{\prime 2} )
          + h_2 \bbox{\chi} \cdot \bbox{\chi}' ],
\end{equation}
with $g=\Omega_M/\Omega_B$ and
\begin{eqnarray}
h_1 &=& \frac{6(12g^2+25g+3)}{25(3+8g)} \Omega_B, \nonumber \\
h_2 &=& \frac{36(4g^2+1)}{25(3+8g)} \Omega_M.
\end{eqnarray}
Finally, using the above informations, we get
\begin{mathletters}
\label{res1}
\begin{eqnarray}
\mid \langle \Lambda K^+ \mid \psi_{5q} (\Lambda K) \rangle \mid^2
&=& \frac{10}{N^2}
\mid 1 -  {\textstyle \frac{1}{12}} \int \d \bbox{\chi} \d \bbox{\chi}'\,
   \varphi_{\Lambda K^+} (\bbox{\chi})
   K ( \bbox{\chi}, \bbox{\chi}' )
   \varphi_{\Lambda K^+} (\bbox{\chi}') \mid^2, \\
\mid \langle p_0 \eta_s \mid \psi_{5q} (\Lambda K) \rangle \mid^2
&=& \frac{10}{N^2}
\mid {\textstyle \frac{1}{4\sqrt6}} \int \d \bbox{\chi} \d \bbox{\chi}'\,
   \varphi_{p_0 \eta_s} (\bbox{\chi})
   K ( \bbox{\chi}, \bbox{\chi}' )
   \varphi_{\Lambda K^+} (\bbox{\chi}') \mid^2, \\
\mid \langle p_0 \phi_s^1 \mid \psi_{5q} (\Lambda K) \rangle \mid^2
&=& \frac{10}{N^2}
\mid {\textstyle \frac{1}{4\sqrt2}} \int \d \bbox{\chi} \d \bbox{\chi}'\,
   \varphi_{p_0 \phi_s} (\bbox{\chi})
   K ( \bbox{\chi}, \bbox{\chi}' )
   \varphi_{\Lambda K^+} (\bbox{\chi}') \mid^2, \\
\mid \langle p_0 \phi_s^3 \mid \psi_{5q} (\Lambda K) \rangle \mid^2
&=& 0.
\end{eqnarray}

For numerical calculations, we use the $\varphi_{BM} (r)$ of the
cloudy bag model, which read
\end{mathletters}
\begin{equation}
\varphi_{BM} (r) = \left\{ \begin{array}{ll}
  \displaystyle
  \frac{C_A}{m_Mr} \left\{ \cosh (m_Mr) - \frac{\sinh (m_Mr)}{m_Mr}
                   \right\} \hskip 1cm   & r \leq R_B, \\
  \displaystyle
  \frac{C_B}{m_Mr} \left( 1 + \frac{1}{m_Mr} \right) e^{-m_Mr}
  & r \geq R_B, \end{array} \right.
\label{meson_wf}
\end{equation}
where $m_M$ is the meson mass and the constants $C_A$,$C_B$ are
fixed by the relations \cite{Myhrer},
\begin{eqnarray}
\frac{C_B}{C_A} &=& \frac12 \left( \frac{b-1}{b+1} e^{2b} + 1 \right),
\nonumber \\
C_A &=& - \frac{b \beta}{m_M} \left( \sinh b + \frac{C_B}{C_A} e^{-b}
\right)^{-1},
\end{eqnarray}
with $b = m_M R_B$ and
\begin{equation}
\beta = \frac{{\cal N}^2}{f_M} j_0 (\Omega) j_1 (\Omega).
\end{equation}
Here, ${\cal N}$ is the normalization constant of the bag, $f_M$ the meson
decay constant, and $\Omega$ the eigenfrequency of the bag with spherical
Bessel functions $j_0$ and $j_1$.

Using the radial wavefunction of (\ref{meson_wf}), we can obtain the
numerical results for the integrals of Eq. (\ref{res1}), which leads to the
probabilities of each cluster in $\psi_{5q}(\Lambda K)$ of (\ref{psi5q}) as
\begin{eqnarray}
{\cal P} (\Lambda K^+) &\approx& 1.0, \nonumber \\
{\cal P} (p_0 \eta_s ) &\approx& 2.7 \times 10^{-7}, \nonumber \\
{\cal P} (p_0 \phi_s^1 ) &\approx& 8.0 \times 10^{-7}, \nonumber \\
{\cal P} (p_0 \phi_s^3 ) &=& 0.0.
\label{prob}
\end{eqnarray}

We also carried out the calculations for other configurations: $\Sigma K$,
$p_0 \eta$, $\Sigma^* K$, and $\Delta \eta$, and obtained
similar results of Eq. (\ref{prob}); i.e., the recoupling effects are
very small. Especially, the probability of $p_0 \phi_s^3$ is zero in
$\psi_{5q} (\Lambda K)$ and in $\psi_{5q} (\Sigma K)$ because of the
spin-flavor structure. By the same reason we have ${\cal P} (p_0
\phi_s^{1,3}) = 0$ in $\psi_{5q} (p \eta)$, ${\cal P} (p_0 \eta_s) =
{\cal P} (p_0 \phi_s^1) = 0$ in $\psi_{5q} (\Sigma^* K)$, and ${\cal
P} (p_0 \eta_s) = {\cal P} (p_0 \phi_s^{1,3}) = 0$ in $\psi_{5q}
(\Delta \eta)$. All these results imply that (i) the strangeness
coefficient $B a_0$ is approximately equal to the amplitude $A_{p_0
\eta}$ of (\ref{p-wf}), of which square is estimated as $\leq 0.5$\%
in the cloudy bag model; (ii) the coefficient $B a_1$ is nearly zero
when we consider only the pseudoscalar meson cloud.
Therefore, to get more accurate value for $Ba_1$, it is necessary to
know the probabilities and radial wavefunctions of the vector meson
cloud. Inclusion of vector meson degrees of freedom in chiral
bag, however, is still incomplete \cite{BVM}.


\section{Summary}

As a summary, we re-estimated the $\phi$ meson electroproduction from
a proton target in $uud$-$s\bar s$ cluster model.  Our calculation is
based on the RHOM which takes into account the main relativistic effect,
namely, the Lorentz contraction of the composite particle wavefunction.

First, we confirmed the results of Ref. \cite{HKW} for the $s\bar s$
knockout process in NRQM. For the knockout cross section to be
comparable to the VDM one, we have to assume $B^2=10 \sim 20$\%, which
is regarded as an upper bound of the strange quark content of the nucleon.
Then we extended the calculation to the study of the $uud$ knockout and
their interference.  We found that the $uud$ knockout and the interference
effect are dominated by the $s\bar s$ knockout and the magnetic part of
the hadron e.m. current gives the main contribution.

Next, we improved the model predictions by taking the RHOM which is
successful in describing nucleon form factors in a wide range of
$\bbox{q}^2$. This successful description comes from the different
functional dependence of the overlap integrals, which can explain the
empirical dipole-type formula of the form factors. In this model we
found that assuming only 3--5\% admixture of the strange quarks in the
nucleon can give a similar size of the cross sections for the
$s\bar s$-knockout and diffractive contributions. The cross sections
of the RHOM are enhanced compared with the NRQM ones by an order of
magnitude. As in NRQM, we could find that the main contribution of the
amplitudes is from the magnetic part of the current. We also found that
the $t$-dependence of the cross section of $uud$ knockout is different
from those of $s\bar s$ knockout and VDM. This strong difference may be
useful in determining the ratio $a_0/a_1$ of Eq. (\ref{protonwf}) and
can be checked in future experiments at CEBAF.

Finally, we examined the strangeness coefficients of the proton
wavefunction in quark model using the cloudy bag model predictions on
the meson clouds. We found that the amplitude $Ba_0$ is nearly equal
to the amplitude of the $\eta$ meson cloud in the proton wavefunction,
which is predicted as less than 0.5\% in the cloudy bag model. We can
not infer the amplitude $Ba_1$ unless we include vector meson clouds
in the model. Nonetheless, this smaller value of $B^2$ may imply an
important role of the VMD diffractive process of the $\phi$
electroproduction. Therefore, to distinguish the knockout and the
diffractive contributions, it will be useful to study other physically
measurable quantities, such as polarization measurements
\cite{HKW,TYO}.


\acknowledgements

We are grateful to Profs. T.-S. H. Lee, M. Namiki and F. Tabakin for
helpful discussions. One of us (A.I.T.) acknowledges
the support from the National Science Council of the Republic of China
as well as the warm hospitality of the Physics Department at
National Taiwan University. This work was supported in
part by the National Science Council of ROC under Grant No.
NSC82-0212-M002-170Y and in part by the ISF Grant No. MP 80000.


\appendix

\section*{}

In this Appendix, we calculate the overlap integrals in RHOM.
Let us consider an overlap integral of the form
\begin{eqnarray}
I = \int \d^4 x \, \e^{ \I q \cdot x} \psi_f(P_f;x) \psi_i(P_i;x),
\end{eqnarray}
where $\psi_{i,f} (P_{i,f};x)$ are the RHOM intrinsic wavefunctions of the
composite two-particle systems with 4-momentum $P_{i,f}$,
\begin{eqnarray}
\psi_{i,f}(P_{i,f},x)=\frac{\Omega_{i,f}}{\pi}
   \exp \Bigl\{ \frac{\Omega_{i,f}}{2} \Bigl[ x^2
   -2 \Bigl( \frac{P_{i,f} \cdot x}{M_{i,f}} \Bigr)^2 \Bigr] \Bigr\} .
\end{eqnarray}
For our purpose we take $\Omega_i = \Omega_f = \Omega$ and $q_\mu$ is an 
arbitrary 4-momentum.
In the laboratory frame, $P_{i,\mu} = (M_i,\bbox{0})$ and
$P_{f,\mu} = (E_f, \bbox{P}_f)$. This integral can be easily evaluated
in the ``$k$-frame," where
\begin{mathletters}
\begin{eqnarray}
\bbox{P}'_f = -\frac{M_f}{M_i}\bbox{P}'_i = \frac{\bbox{k}}{2},
\end{eqnarray}
with
\begin{equation}
\bbox{k}^2 = 2 M_f^2 \left( \frac{E_f}{M_f} - 1 \right).
\end{equation}
\end{mathletters}

Then in the $k$-frame the integration can be carried out straightforwardly
and the overlap integral becomes
\begin{eqnarray}
I = ( 1 + \frac{\bbox{k}^2}{2M_f} )^{-1}
    \exp \{ -\frac{1}{4\Omega} [\bbox{{q'}}^2_{\perp}+
    ({{q'}^2_0+\bbox{{q'}}^2_{\parallel}})
    (1+\frac{\bbox{k}^2}{2M_f})^{-1} ] \},
\end{eqnarray}
where $\bbox{{q'}}_{\perp}$ and $\bbox{{q'}}_{\parallel}$ are the components
of $\bf q'$ in the $k$-frame, which are perpendicular and parallel to
$\bf k$, respectively. Transformation to the laboratory frame gives
\begin{eqnarray}
I = \frac{M_f}{E_f} \exp(-\frac{1}{4\Omega}q_{\rm eff}^2),
\end{eqnarray}
where
\begin{eqnarray}
q^2_{\rm eff} =
 \bbox{q}^2+q^2_0-2\frac{q_0}{E_f}\bbox{q} \cdot \bbox{P}_f.
\end{eqnarray}

Extending this result to an $n$-particle system leads us to
\begin{eqnarray}
I_n = \left(\frac{M_f}{E_f}\right)^{n-1} \exp(-\frac{1}{4\Omega}
    q_{\rm eff}^2),
\label{OIn}
\end{eqnarray}
which is used to calculate the overlap integrals in Sec. V.
In the {\it elastic\/} scattering limit, where $M_i=M_f=M$,
$q_\mu=A(P_f - P_i)_\mu$ with a constant $A$, then $E_f/M=1+Q^2/(2M^2)$
and Eq. (\ref{OIn}) is reduced to the well-known result \cite{FKN}
\begin{eqnarray}
I_n = ( 1+\frac{Q^2}{2M^2} )^{1-n}
\exp(-\frac{1}{4\Omega}q_{\rm eff}^2), \qquad
q^2_{\rm eff} =
 A^2 Q^2 ( 1 + \frac{Q^2}{2M^2} )^{-1},
\end{eqnarray}
with $Q^2 = - (P_f - P_i)^2$.



\vskip 2.5cm

\vskip 2.5cm

\end{document}